\documentclass[a4paper,11pt]{article}
\pdfoutput=1
\usepackage{jinstpub}
\usepackage{graphicx, amsthm, multirow}
\usepackage{mathrsfs}

\title{Topological track reconstruction in unsegmented, large-volume liquid scintillator detectors}
\author[a]{Bj\"orn S. Wonsak,}
\author[a]{Caren I. Hagner,}
\author[b]{Dominikus A. Hellgartner,}
\author[c]{Kai Loo,}
\author[d,*]{Sebastian Lorenz,}
\author[a]{David J. Meyh\"ofer,}
\author[b]{Lothar Oberauer,}
\author[a]{Henning Rebber,}
\author[c]{Wladyslaw H. Trzaska}
\author[d]{and Michael Wurm}
\affiliation[a]{University of Hamburg, Institute of Experimental Physics, Luruper Chaussee 149, 22761 Hamburg, Germany}
\affiliation[b]{Technical University Munich, Department of Physics, James-Franck-Str.\,1, 85748 Garching at Munich, Germany}
\affiliation[c]{University of Jyvaskyla, Department of Physics, Survontie 9 C, FI-40014 Jyvaskyla, Finland}
\affiliation[d]{Johannes Gutenberg-University Mainz, Institute of Physics, Staudingerweg 7, 55128 Mainz, Germany}
\emailAdd{slorenz@uni-mainz.de}

\abstract{Unsegmented, large-volume liquid scintillator (LS) neutrino detectors have proven to be a key
technology for low-energy neutrino physics. The efficient rejection of radionuclide background induced by
cosmic muon interactions is of paramount importance for their success in high-precision MeV neutrino 
measurements. We present a novel technique to reconstruct GeV particle tracks in LS, whose main
property, the resolution of topological features and changes in the differential energy loss 
$\mathrm{d}E/\mathrm{d}x$, allows for improved rejection strategies. Different to common track 
reconstruction approaches, our method does not rely on concrete track / topology hypotheses. 
Instead, based on a reference point in space and time, the observed distribution of 
photon arrival times at the photosensors and the detector's characteristics in terms of photon production,
propagation and detection (optical model), it reconstructs the voxelized distribution of optical photon
emissions. Techniques from three-dimensional data analysis can then be applied to extract parameters
describing the topology, e.g., the direction of a track. We performed a first performance evaluation 
of our method using single muon events with up to 10\,GeV from a Geant4 simulation of the LENA detector. 
The current results indicate that our approach is competitive with existing reconstruction methods -- 
although its full potential has not yet been exploited. We also remark on other detector technologies 
in astroparticle physics as well as applications in medical imaging that could benefit from the 
fundamental ideas of our method.}


\arxivnumber{1803.08802} 

\begin{document}
\maketitle
\flushbottom


\section{Motivation}
\label{sec:Motivation}

Over the last decades, unsegmented liquid scintillator (LS) detectors have contributed substantially to 
neutrino physics: The KamLAND measurement of reactor antineutrino disappearance was one of the 
keystones for unraveling neutrino flavor oscillations~\cite{Eguchi:2002_KL_FirstRes}. Borexino achieved 
the first spectroscopic measurement of sub-MeV neutrinos from the Sun and confirmed the MSW-LMA 
scenario for solar neutrino oscillations \cite{Arpesella:2008_BX_Be7, Bellini:2008_BX_B8, 
Collaboration:2011_BX_pep}. Finally, the three km-baseline reactor experiments Daya Bay, RENO and 
Double Chooz have performed a high-precision measurement of the small mixing angle 
$\theta_{13}$~\cite{An:2012_DYB_NuEDisapp, Ahn:2012_RENO_NuEDisapp, Abe:2012_DC_NuEDisapp}. 

All these experiments were aiming at a neutrino energy range from several hundred keV to a 
few tens of MeV. At these energies, the isotropic emission of scintillation light allows only for a 
point-like reconstruction of the neutrino events, returning no directional information. However, it 
has been shown in~\cite{Learned:2009_HeNuIntInLSc, Peltoniemi:2009_LScHeTracking} that this 
apparent limitation of the technique is lifted for higher neutrino energies, as soon as the track 
length of the final state particles exceeds a few tens of centimeters.\footnote{Note that there are
efforts ongoing~\cite{Caravaca:2016_CHESS, Alonso:2014_ASDC, Beacom:2017_Jinping} to use a combination of more transparent 
scintillators with delayed light emission and/or faster light sensors to resolve the directional 
Cherenkov cone preceding the scintillation light front, aiming for track reconstruction at MeV energies.}

Moreover, almost all of the above mentioned experiments employ algorithms for the reconstruction of 
the extended tracks of cosmic muons (e.g., see~\cite{Bellini:2011_BX_MuonNeutDet}). Precise 
reconstruction is important in order to apply spatial vetoes against background from cosmogenic 
radionuclides~\cite{Collaboration:2011_BX_pep, Abe:2012_DC_NuEDisapp}. Corresponding analyses have
demonstrated that radioisotope production is strongly correlated with muon-induced 
showers~\cite{Abe:2009_KL_Cosmogenics}. Therefore, more efficient cosmogenic background rejection 
strategies are within reach if spatial vetoes can be focused on local increases in energy 
deposition from such showers. Especially huge LS detectors, like JUNO, are prone 
to high rates of muon-induced background~\cite{An:2015_JUNO, Grassi:2014_ShoweringCosmogenics, 
Grassi:2015_CosmogenicVetoing}; all the more so 
if the experiment is shielded against muons only by a shallow overburden.

Directional information or an even more detailed reconstruction of final-state particle tracks from 
GeV neutrino events are highly desirable for the next generation of large-volume neutrino 
detectors. This is especially true in the context of long-baseline oscillation experiments aiming 
at the determination of the neutrino mass ordering and the search for CP violation, either 
with accelerator-based neutrino beams or the oscillations of atmospheric neutrinos 
\cite{Wurm:2011_LENA}. Moreover, track reconstruction of the decay particles would enhance the 
sensitivity of LS detectors in proton decay searches, possibly expanding the number of accessible channels beyond the 
golden channel $p\to K^+\overline{\nu}$~\cite{Undagoitia:2005_LENA_PDecay}. A precise muon track 
reconstruction can help to discriminate the muonic $K^+$ decay from background induced by atmospheric
neutrinos.

This paper presents a new method for the reconstruction of GeV particle tracks in LS, which was 
developed in the context of the LENA and LAGUNA-LBNO design studies~\cite{Wurm:2011_LENA}. 
The algorithm is sufficiently variable to be applied to differing detector geometries, making it of 
equal interest to running and near-future LS experiments, e.g., Borexino, SNO$+$ and 
JUNO~\cite{Bellini:2011_BX_MuonNeutDet, Chen:2005_SNOplus, An:2015_JUNO}. Beyond earlier 
findings~\cite{Learned:2009_HeNuIntInLSc, Peltoniemi:2009_LScHeTracking}, we verify the performance 
of the reconstruction algorithm with muon tracks from a state-of-the-art 
Geant4~\cite{Agostinelli:2002_Geant4Pub1, Allison:2006_Geant4Pub2} simulation of the LENA detector.

In section \ref{sec:EventsInLsc}, we give a general overview of how an event in a LS detector 
generates observable information in the form of scintillation light, how this information can be 
used to reconstruct charged particle tracks and which effects need to be taken into account for 
this purpose. The presentation of the new, topological reconstruction method follows in 
section \ref{sec:Method} as the focus of this paper. Our first analysis of the method's performance 
based on the simulated muons in LENA is detailed in section \ref{sec:Analysis}. After an 
outlook on possible further advancements and applications of the presented reconstruction technique
in section \ref{sec:Outlook}, we conclude with a summary.

\section{Events in liquid scintillator detectors}
\label{sec:EventsInLsc}

Liquid scintillator becomes excited and emits light when a charged particle loses kinetic energy
while traversing the medium. The number of emitted scintillation photons per unit length
varies along the path of the particle and depends on the differential energy loss $\mathrm{d}E/
\mathrm{d}x$. Since
the light is emitted in several competing transitions, the probability density function (PDF) $\Phi_{\textrm{em}}(t)$ of 
the fluorescence time profile can be approximated by a weighted sum of exponential decay functions 
with different decay time constants $\boldsymbol\tau = \{ \tau_{1},
\dots, \tau_{n}\}$ and weights $\boldsymbol\omega = \{ \omega_{1},\dots, 
\omega_{n}\}$~\cite{Birks:1964_TheoPracScintCount}:
\begin{equation}
  \Phi_{\textrm{em}}(t;\boldsymbol{\tau},\boldsymbol\omega) = \sum_{i=1}^{n}
  \frac{\omega_{i}}{\tau_{i}} e^{-\frac{t-t_{0}}{\tau_{i}}} \, , \quad 
  t \geq t_{0} \, , \quad
  \sum_{i=1}^{n} \omega_{i} = 1 \, .
  \label{eq:ScintEmissionTimingPDF}
\end{equation}
Additionally, there is a fast rising flank ($\lesssim 1$\,ns~\cite{Caravaca:2016_CHESS}) of the 
time profile that depends on the concentration of wavelength shifter in 
the LS. The parameter $t_0$ is the point in time of the excitation.

\begin{figure}[b!t]
  \centering
    \includegraphics[width=0.70\textwidth]{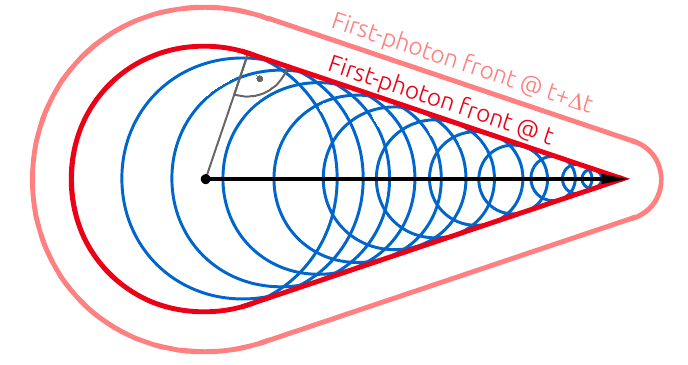}
    \caption{
Illustration of the scintillation photon emission along a straight track (black) in LS from a 
charged particle with speed $v = \beta c_0$. Cherenkov light production~\cite{Olive:2016_PDG} is not 
taken into account. Fast, spherical emissions of scintillation light (blue)
along the particle trajectory superimpose each other and form a cone-like structure, the 
first-photon front. Contrary to the Cherenkov case, the cone tip becomes more and 
more rounded with further time $\Delta t$ for photon propagation (light red) after the stop of the 
particle at time $t$ (dark red). Moreover, there is a significant emission of light into 
the half-space lying opposite to the track direction.}
  \label{fig:ScintTrackSchematic}
\end{figure}

As illustrated in figure~\ref{fig:ScintTrackSchematic}, scintillation photons are emitted 
isotropically from each point of the charged particle's track and propagate with 
the speed of light in the medium, i.e., the group velocity $v_g(\lambda)$ dependent of the photon 
wavelength $\lambda$.

In addition to scintillation light, Cherenkov photons are emitted if the charged particle's speed exceeds the local phase 
velocity of light in the medium~\cite{Olive:2016_PDG}. Their relative contribution to 
the total emitted light per unit track length depends on the particle's speed and ionization energy 
loss as well as the scintillation light yield of the LS. For a LAB-based LS mixture with additions
of wavelength-shifting PPO and Bis-MSB as proposed for LENA~\cite{Wurm:2011_LENA}, giving a 
scintillation light yield of about $10^4/\mathrm{MeV}$, the Cherenkov light contribution typically is 
$\lesssim7\%$. However, it can be significantly higher for other LS mixtures.
Contrary to the delayed isotropic emission of scintillation photons, Cherenkov radiation is emitted 
instantaneously, forming a conical light front centered on the particle track (e.g., see 
\cite{Olive:2016_PDG}).

The reconstruction of event vertices, i.e., particle energies and spatial topologies, relies on the 
amount and timing of photons detected by photosensors surrounding the LS volume. Large area photomultiplier tubes 
(PMTs), in many cases armed with light concentrators (LCs), are the option commonly chosen. 
Depending on the read-out electronics, an ADC\,/\,TDC scheme allows to determine the arrival time 
of only the first photon at a given PMT (``first hit''). Alternatively, FADC read-out records
the entire time structure of all photon hits (``full pulse shape''). 
While point-like vertices and simple single-track events can be reconstructed 
based on the former, full pulse shape information is essential for determining more complex 
topologies. 

In case single particles deposit some hundreds of MeV or more in the scintillator, the region of 
energy deposition is spatially extended over several meters distance. In contrast to Cherenkov 
emission, the isotropic scintillation light bears no directional information on a photon-to-photon 
basis. However, the charged particle passes through the scintillator with almost the speed of light in
vacuum, $c_0$, whereas the scintillation photons only propagate with $v_g(\lambda)$. Due to this, 
the scintillation photons are enclosed in a cone-like volume that surrounds the particle track 
(cf. figure~\ref{fig:ScintTrackSchematic} and, e.g., figure 14 in~\cite{Wurm:2011_LENA}) while the particle
keeps moving through the scintillator.
Especially the surface defined by the front of first photons emitted from each point on the 
particle's path has a characteristic cone-like shape. It is similar to the Cherenkov case in 
forward direction, but features an additional backward running spherical light front. If the 
particle stops inside the scintillator, a similar spherical front develops in forward direction. Since 
the cone shape is projected onto the PMTs over time, its structure and orientation is encoded in the photon hit 
time distribution of first hits~\cite{Learned:2009_HeNuIntInLSc}. Behind the first-photon front, 
a ``scintillation photon field'' populated via delayed emissions 
(see eq.~(\ref{eq:ScintEmissionTimingPDF})) and photon scattering fills the cone-like volume. 
For a given point inside the cone, the local density of scintillation photons decreases over time due to
the field's dispersion and photon absorption. Significant, localized increases in $\mathrm{d}E / \mathrm{d}x$,
e.g., from increased ionization by the stopping primary particle or the generation of secondary particle
tracks, seed locally higher photon densities and populations of the first-photon front. These features
dissolve over time when they spread out until they reach the light sensors. In principle,
variations in the population of the first-photon front from topological peculiarities would be reflected
in the collected charge pattern at the PMTs that is obtained with a fixed, short collection time after
each PMT's first hit and thus is sensitive to multi-photon and/or fast consecutive hits. However, the 
contrast and details of the reconstructed topology increases in
case the change of the photon density over time is probed by recording full PMT pulse shapes. 
The inference from the observed pulse shapes to the topology (features) of the event
is the purpose of the reconstruction method described in the next section.

The reconstruction quality critically depends on two properties of the scintillation light 
emission: First, for a given track segment, there must be a strong correlation between the 
point in time of the energy deposition and the subsequent photon emission. Since the scintillation
photon emission is a stochastic process, described by the timing PDF in 
eq.~(\ref{eq:ScintEmissionTimingPDF}), fast scintillators with short decay time constants will 
better preserve this correlation. Second, the quality of the original time 
information carried by all the 
emitted photons is maintained as much as possible. This makes the transparency of the LS an 
important factor, especially for low-energy events. It depends on the individual impacts of
photon absorption and scattering~\cite{Wurm:2010_OptScat}. Compared to the ideal case of linear photon 
trajectories, scattering of photons randomly changes their flight paths -- and thus their times of flight -- until detection or absorption. Since photon 
trajectories cannot be reconstructed, scattering will distort the observed photon hit time 
distribution, i.e., the basis for the reconstruction.

\section{Method of topological reconstruction}
\label{sec:Method}

Track reconstruction algorithms based on linear track hypotheses for muons in LS have been 
successfully employed in Borexino and KamLAND and other experiments~\cite{ 
Bellini:2011_BX_MuonNeutDet, Abe:2009_KL_Cosmogenics}. In this section, we present a novel, 
iterative reconstruction method that returns a three-dimensional density distribution of 
scintillation photon emissions.\footnote{The idea 
of the reconstruction method can in principle be adapted to pure Cherenkov light or the more 
complex hybrid case, dominant scintillation light with some Cherenkov light contribution, relevant 
for LS detectors. For simplicity, we only describe the case of pure scintillation light in this 
paper.} This distribution is a direct, topological representation of 
the event's energy deposition. Contrary to algorithms assuming a given number of linear tracks, 
this method is free of any hypothesis concerning the event type or, more importantly, the 
geometrical appearance of the energy deposition. The only assumption made for the topological 
reconstruction is that a reference point on the topology is known reasonably well in space and time 
(the time of the energy deposition) and that the particle going through this point moves on a 
straight line with speed $c_0$. For example, the reference point could be provided by an external 
muon tracking system or by other reconstruction methods based on timing information, like the 
``backtracking-algorithm''~\cite{Hellgartner:2015_PhDThesis}, or geometrical 
models~\cite{Abe:2014_DCmuReco, Genster:2018_ConeMethod}.

In section \ref{sec:Method.BasicIdea}, we describe the basic idea of the new reconstruction method, 
i.e., of one iteration (see below). For simplicity, the description is limited to the case of a single track.
To demonstrate the method's ability to provide information not only on the locations of 
energy depositions but also on the spatial energy distribution, we assume that full pulse shape
information is available.\footnote{For experiments in which only the first photon hit time 
at each PMT is available, this method can, after some adaptations, still be useful to provide pure 
topological information. However, the information on the energy deposition in each point is largely 
lost in this case.} Section \ref{sec:Method.Iteration} introduces the concept of a ``probability 
mask'' and explains how it turns the topological reconstruction into an iterative procedure.

\subsection{Basic reconstruction strategy}
\label{sec:Method.BasicIdea}

Our general approach to track reconstruction in LS is to use the timing information 
provided by individual PMT hits to construct three-dimensional PDFs for the origins
of the detected scintillation photons. The PDFs are superimposed to emphasize the spatial regions 
where scintillation photons have likely been emitted, i.e., locations of energy depositions along 
the searched-for particle track.

\begin{figure}[b!t]
  \centering
    \includegraphics[width=0.70\textwidth]{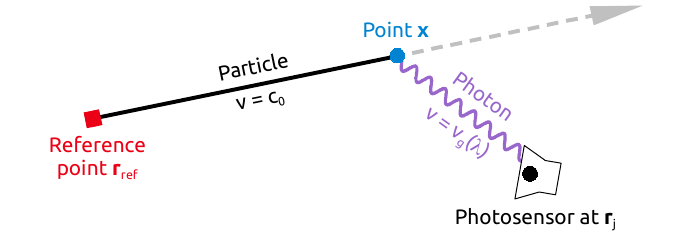}
    \caption{
Illustration of the basic part of the information transport model used in the 
topological reconstruction. Passing the reference point $\mathbf{r}_{\mathrm{ref}}$ at the
reference time $t_{\mathrm{ref}}$, a charged particle travels with the speed of light in vacuum
$c_0$ along a straight track through $\mathbf{x}$. A photon emitted with wavelength 
$\lambda$ at point $\mathbf{x}$ reaches the photosensor at $\mathbf{r}_j$ with the speed equal to 
the wavelength-dependent group velocity $v_g(\lambda)$.}
  \label{fig:InfoTransportModel}
\end{figure}

For a point-like source of scintillation photons and perfect time information, the possible origins 
of the detected photons define spherical isochrones around the PMTs, which overlap at the position 
of the source. For an extended event formed by a charged particle track, however, this approach has 
to be modified because a given photon can potentially originate from any point along the track. 
The non-zero time of flight of the particle acts as an offset to the photon emission 
time. As a consequence, we need a model for the transport of the information ``\emph{the charged 
particle caused scintillation photon emission in point $\mathbf{x}$ and the subsequent detection of 
a hit at the $j$th PMT}'' from the unknown origin ($\mathbf{x}$) to the $j$th PMT ($\mathbf{r}_j$). 
An illustration of our basic model is shown in figure \ref{fig:InfoTransportModel}.

It contains two assumptions: i) a reference point in space ($\mathbf{r}_{\mathrm{ref}}$) and time 
($t_{\mathrm{ref}}$) on the track is known and ii) the charged particle moves with the speed of 
light in vacuum $c_0$ along a straight line through this reference point. Based on this model, the 
theoretical arrival time $\hat{t}_j(\mathbf{x})$ at a PMT at $\mathbf{r}_j$ of a photon emitted 
from point $\mathbf{x}$ can be calculated as 
\begin{equation}
\hat{t}_j(\mathbf{x}) \equiv
\hat{t}_j(\mathbf{x}; \mathbf{r}_j, \mathbf{r}_{\mathrm{ref}}, t_{\mathrm{ref}}) = t_{\mathrm{ref}} 
\pm 
    \underbrace{\frac{|\mathbf{x} -\mathbf{r}_{\mathrm{ref}}|}{c_0}}_{\mathrm{particle}}
    + \underbrace{\vphantom{\frac{|\mathbf{x} -\mathbf{r}_{\mathrm{ref}}|}{c_0}}
    t_{\mathrm{ph}}(\mathbf{x},\mathbf{r}_j)}_{\mathrm{photon}} + t_s \, .
\label{eq:RecoTimeModel}
\end{equation}
The sign of the time contribution from the particle depends on whether 
point $\mathbf{x}$ was reached in time before $(-)$ or after $(+)$ the reference point 
$\mathbf{r}_{\mathrm{ref}}$. Due to the straight track, the particle contribution to 
$\hat{t}_j(\mathbf{x})$ is easy to calculate. The photon, however, can take slightly different paths 
from the emission point $\mathbf{x}$ to a point on the photosensitive area of the PMT at $\mathbf{r}_j$. 
The random variable 
$t_{\mathrm{ph}}(\mathbf{x},\mathbf{r}_j)$ for the photon contribution therefore has a PDF 
$\Phi_{t_{\mathrm{ph}}}(t;\mathbf{x},\mathbf{r}_j)$, which effectively includes the optical model of 
the LS detector (i.e., photon emission, propagation and detection). Finally, the last addend 
$t_s$ accounts for remaining random time contributions, e.g., from delayed scintillation photon 
emission and the transit time jitter of the photosensor, and has the PDF $\Phi_{t_s}(t)$.

\begin{figure}[b!t]
  \centering
  \begin{minipage}{0.48\textwidth}
    \includegraphics[trim=0.1cm 0.1cm 0.5cm 0.1cm,clip=true,width=\textwidth]
      {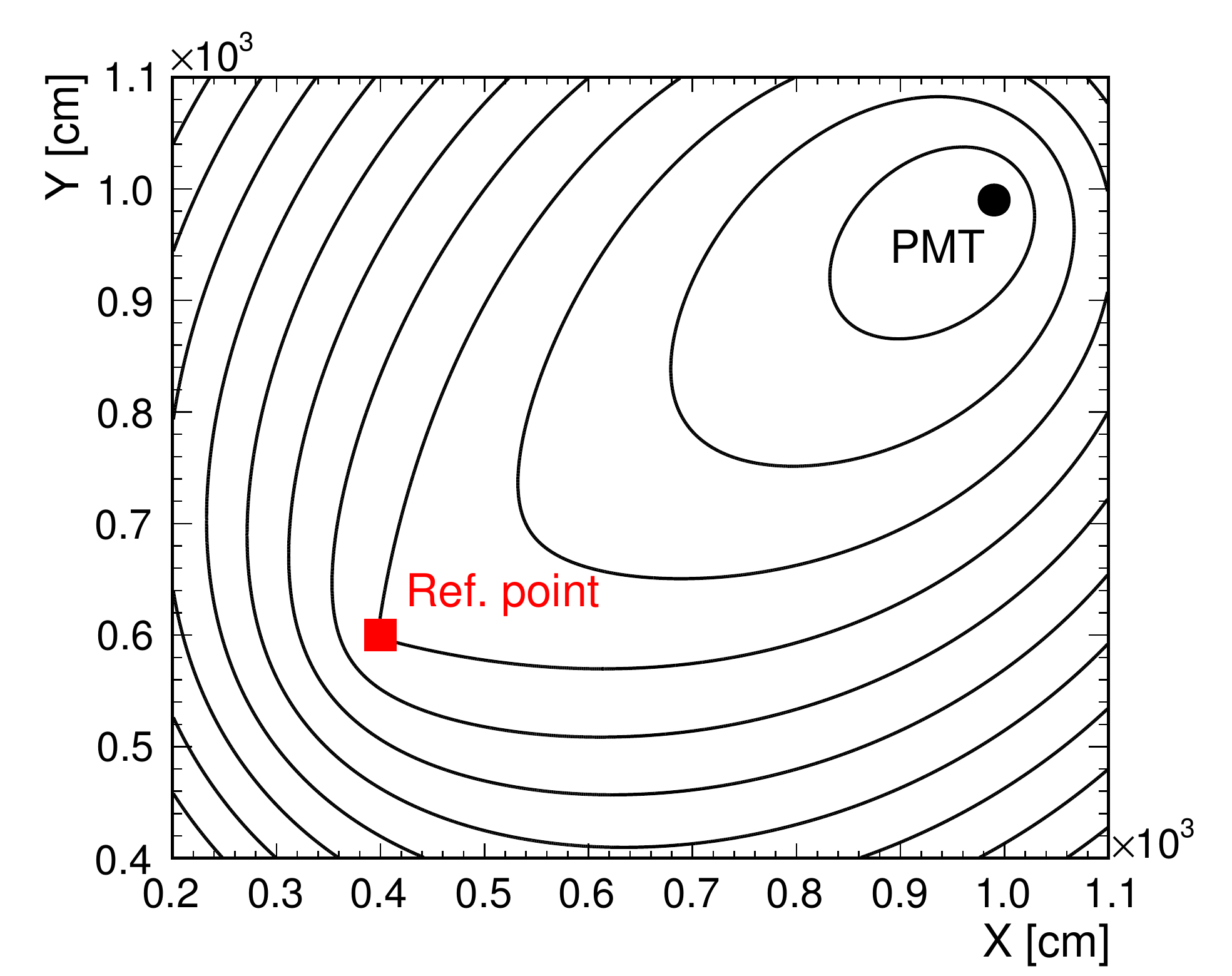}
  \end{minipage}
  ~
  \begin{minipage}{0.48\textwidth}
    \includegraphics[trim=0.1cm 0.1cm 0.5cm 0.1cm,clip=true,width=\textwidth]
      {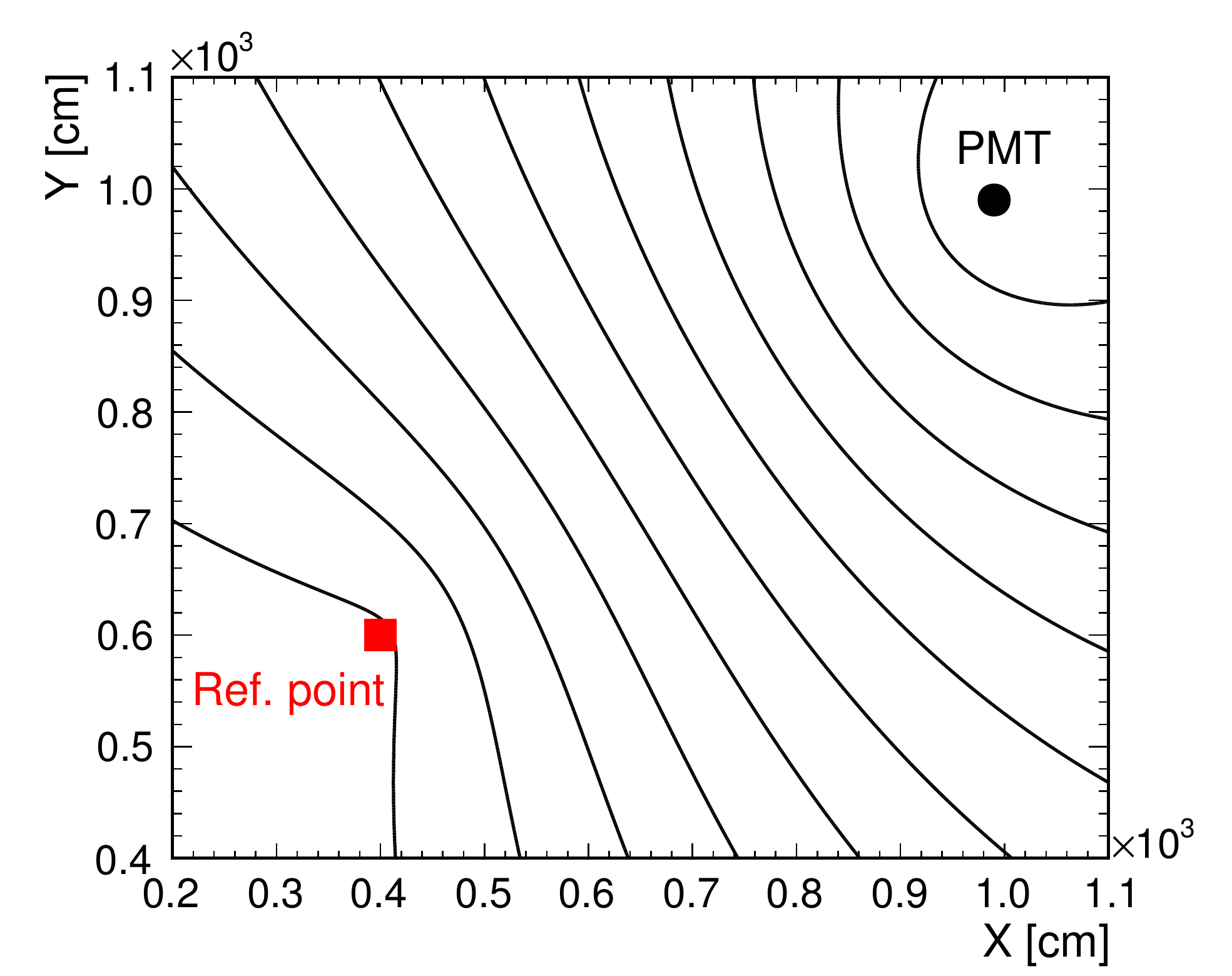}
  \end{minipage}
  \caption{
Two-dimensional isochrones for different values of $\hat{t}_j(\mathbf{x})$ from 
eq.~(\ref{eq:RecoTimeModel}) with the plus (left) or minus (right) sign for the term 
describing the particle contribution. In the computation of $\hat{t}_j(\mathbf{x})$ relative to the 
reference point (red square) at $(400, 600)$ cm, $t_s$ was set to zero and 
$t_{\mathrm{ph}}(\mathbf{x},\mathbf{r}_j)$ was calculated as the time of flight for a direct path 
from $\mathbf{x}$ to the photosensor (black circle) located at $(990, 990)$ cm. For simplicity, 
the phase velocity $v=c_0 / n$ with the refractive index $n = 1.484$ was used for the photon.}
  \label{fig:DropShapes}
\end{figure}

For given values of $t_{\mathrm{ph}}(\mathbf{x},\mathbf{r}_j)$ and $t_s$ in 
eq.~(\ref{eq:RecoTimeModel}), photon emissions from more than one point $\mathbf{x}$ can result
in the same theoretically expected arrival time $\hat{t}_j(\mathbf{x})$. 
The possible origins $\mathbf{x}$ form three-dimensional isochrones in space. 
This is illustrated in figure \ref{fig:DropShapes} for a two-dimensional plane containing the reference 
point and the PMT. The resulting three-dimensional surfaces are rotationally symmetric with 
respect to the axis defined by these two points. This yields a drop-like shape in the 
case of the positive sign for the particle contribution (figure \ref{fig:DropShapes} left) 
and the drop-like equivalents of hyperbolas for the negative sign (figure \ref{fig:DropShapes} 
right). 

\begin{figure}[b!t]
  \centering
  \begin{minipage}{0.48\textwidth}     
    \includegraphics[trim=0.1cm 0.1cm 0.0cm 0.1cm,clip=true,width=\textwidth]
      {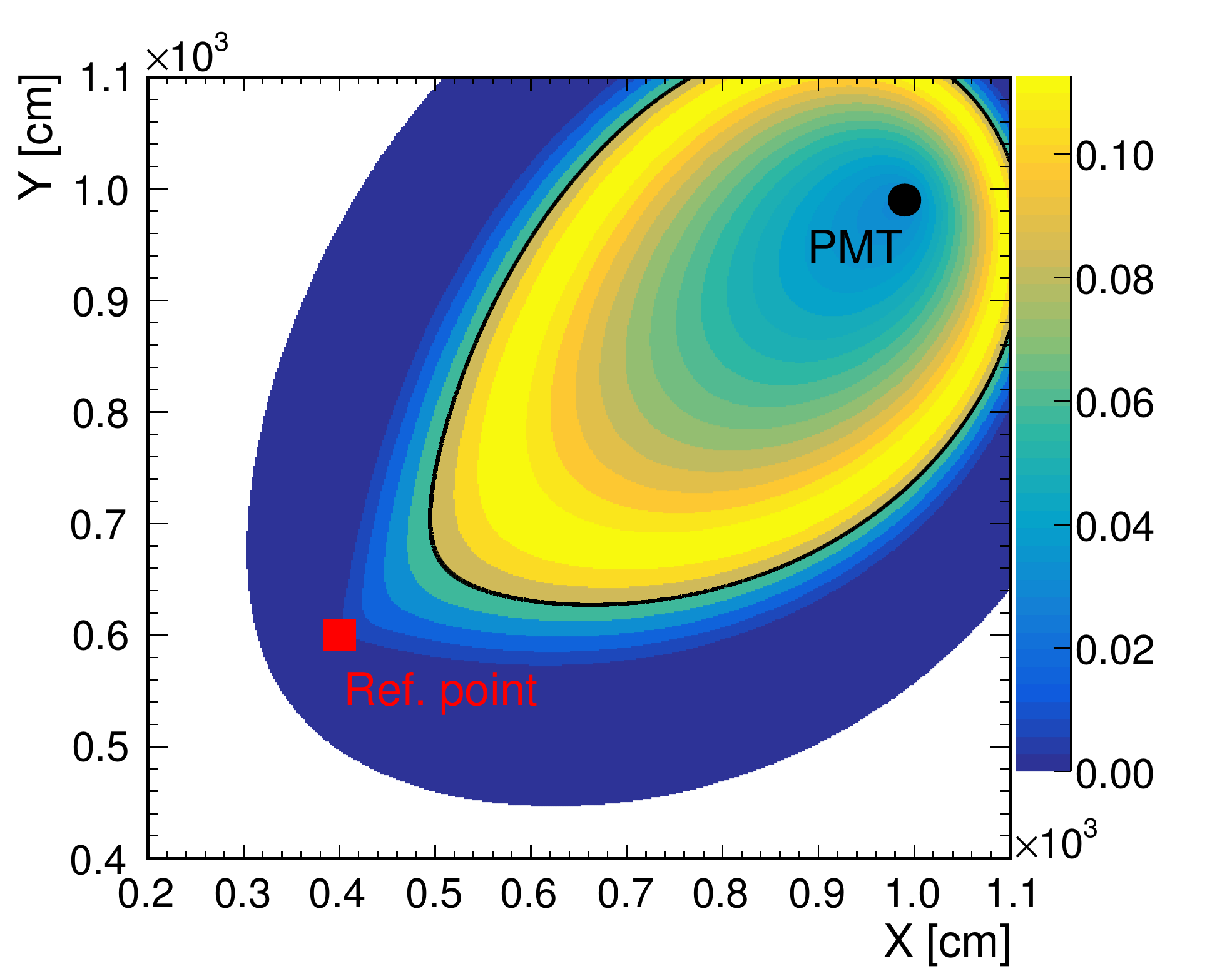}
  \end{minipage}
  ~
  \begin{minipage}{0.48\textwidth}
    \includegraphics[trim=0.1cm 0.1cm 0.0cm 0.1cm,clip=true,width=\textwidth]
      {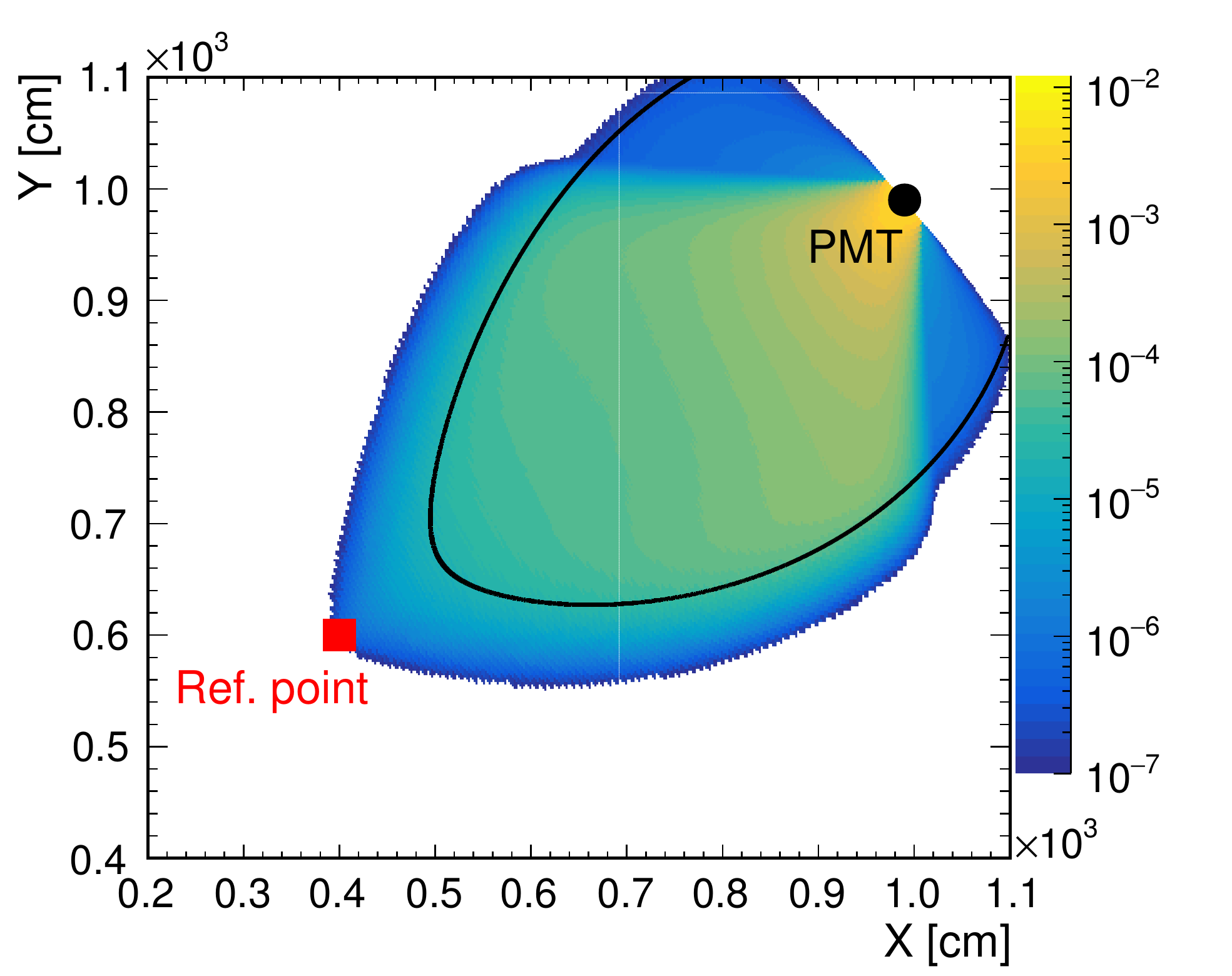}
  \end{minipage}
  \caption{
Unnormalized, two-dimensional versions of $\Phi_{j,k}(\mathbf{x})$ with (right) and without (left) 
the inclusion of the spatial detection efficiency $\varepsilon_j(\mathbf{x})$ for the same configuration 
of photosensor and reference point as in figure~\ref{fig:DropShapes}. The photon hit time for the 
sharp isochrone (black line), calculated as in figure~\ref{fig:DropShapes} with the plus sign for 
the particle contribution, was set to 33\,ns.}
  \label{fig:SmearedDropShape}
\end{figure}

Based on eq.~(\ref{eq:RecoTimeModel}) and the PDFs 
$\Phi_{t_{\mathrm{ph}}}(t;\mathbf{x},\mathbf{r}_j)$ and $\Phi_{t_s}(t)$, the basic procedure of the 
topological reconstruction is to compute the probability density
\begin{equation}
 \Phi_{j,k}(\mathbf{x}) = w_{j,k} \, \varepsilon_j(\mathbf{x}) \int\limits_0^\infty
 \Phi_{t_s}(\Delta t) \,
 \Phi_{t_{\mathrm{ph}}}(t';\mathbf{x},\mathbf{r}_j) \, \mathrm{d} t'
 \label{eq:SinglePhotPDF}
\end{equation}
at every point $\mathbf{x}$ in the LS-filled volume $V_{\mathrm{LS}}$ for a given
combination of $\mathbf{r}_{\mathrm{ref}}$, $\mathbf{r}_j$ and $t_{\mathrm{ref}}$ and
each \underline{individual} photon hit time $t_{j,k}$, where the index pair $j,k$ 
denotes the $k$th photon hit at the $j$th photosensor. It depends on the 
chance that $t_{\mathrm{ph}}(\mathbf{x},\mathbf{r}_j)$ and $t_s$ in eq.~(\ref{eq:RecoTimeModel}) 
have fitting values to get a match between the observed hit time and the hit time predicted by the 
model, $\hat{t}_j(\mathbf{x}) = t_{j,k}$. This is expressed by 
$\Phi_{t_s}(\Delta t)$ with $\Delta t = t_{j,k} - \hat{t}_j(\mathbf{x}; t_{\mathrm{ph}} = t', t_s = 
0)$. It calculates the probability density for the case that the random timing uncertainty $t_s$ 
accounts for the time difference $\Delta t$ between the observed photon hit time $t_{j,k}$ and the
hit time $\hat{t}_j(\mathbf{x}; t_{\mathrm{ph}} = t', t_s = 0)$ expected solely from the flight 
time of the particle and any given photon flight time $t_{\mathrm{ph}} = t'$. Different realizations
of the photon's path through the scintillator medium onto the extended photosensitive surface of 
the sensor are weighted by $\Phi_{t_{\mathrm{ph}}}(t';\mathbf{x},\mathbf{r}_j)$. The spatial 
detection efficiency $\varepsilon_j(\mathbf{x})$ accounts for the fact that the detection of a photon 
by the $j$th sensor depends on spatial constraints (like the solid angle of the photosensor area, angular 
acceptance of a potential LC, photon scattering and photon absorption). The combination of
$\varepsilon_j(\mathbf{x})$ and $\Phi_{t_{\mathrm{ph}}}(t;\mathbf{x},\mathbf{r}_j)$ effectively 
includes the optical model of the LS detector. The factor $w_{j,k}$ in eq.~(\ref{eq:SinglePhotPDF}), 
\begin{equation}
  w_{j,k}^{-1} = \int\limits_{V_{\mathrm{LS}}}
  \varepsilon_j(\mathbf{x}) \int\limits_0^\infty
  \Phi_{t_s}(\Delta t) \,
  \Phi_{t_{\mathrm{ph}}}(t';\mathbf{x},\mathbf{r}_j) \, \mathrm{d} t' \mathrm{d} V \, ,
  \label{eq:NormFactor}
\end{equation}
ensures the proper normalization for a single photon,
\begin{equation}
 \int\limits_{V_{\mathrm{LS}}} \Phi_{j,k}(\mathbf{x}) \mathrm{d} V \overset{!}{=} 1 \, .
 \label{eq:SinglePhotonNorm}
\end{equation}
An example for an unnormalized two-dimensional version of $\Phi_{j,k}(\mathbf{x})$ is shown in 
figure \ref{fig:SmearedDropShape} with (right) and without (left) the effects of the spatial detection
efficiency $\varepsilon_j(\mathbf{x})$. In the left panel, the random time contribution $t_s$ 
leads to a smearing of the single photon PDF perpendicular to the initially sharp isochrone. If 
$\varepsilon_j(\mathbf{x})$ is included (right panel), this efficiency
distribution dominates the picture. The photosensor's field of view, defined by the angular 
acceptance of the considered LC, and the boundary of the detector are clearly visible.

The PDF $\Phi_{j,k}(\mathbf{x})$ is a spatial probability density distribution 
for the emission point of a single photon detected by a single PMT. Therefore, the \emph{spatial density 
distribution corresponding to the origins of all \underline{detected} photons}, 
$\Gamma_{\mathrm{det}}(\mathbf{x})$, can be reconstructed from the superposition of all 
$\Phi_{j,k}(\mathbf{x})$:
\begin{equation}
\Gamma_{\mathrm{det}}(\mathbf{x}) =
\sum \limits_{j} \Gamma_{\mathrm{det},j}(\mathbf{x}) = 
\sum \limits_{j,k} \Phi_{j,k}(\mathbf{x})\, .
\label{eq:NumDensDistDetect}
\end{equation}
In order to get the \emph{spatial density distribution of} \underline{\emph{all} (i.e., detected and 
undetected)} \emph{scintillation photon emissions} $\Gamma_{\mathrm{em}}(\mathbf{x})$, one has to rescale
$\Gamma_{\mathrm{det}}(\mathbf{x})$ by the global detection efficiency $\varepsilon(\mathbf{x})$ according
to 
\begin{equation}
\Gamma_{\mathrm{em}}(\mathbf{x}) =
\frac{\Gamma_{\mathrm{det}}(\mathbf{x})}{\varepsilon(\mathbf{x})} \, .
\label{eq:NumDensDistEmitted}
\end{equation}
The global detection efficiency $\varepsilon(\mathbf{x})$ follows from the single detection efficiencies as
\begin{equation}
\varepsilon(\mathbf{x}) = \sum \limits_{l}\varepsilon_l(\mathbf{x})\, ,
\label{eq:LocalDefEff}
\end{equation}
where the sum with index $l$ includes all active photosensors with and without photon hits.

\subsection{Iterating with a probability mask}
\label{sec:Method.Iteration} 

So far, the reconstruction result as in eqs.~(\ref{eq:NumDensDistDetect}) or 
(\ref{eq:NumDensDistEmitted}) was obtained by summing the PDFs $\Phi_{j,k}(\mathbf{x})$ for all 
available signal photons. From the point of view of statistics this is the correct way if all 
signals can be regarded as independent. In contrast, if the signals are directly 
correlated, as it is the case for a point-like event, the individual, three-dimensional PDFs 
are multiplied to identify the most likely common point of origin in the most efficient way.
However, the latter approach removes the possibility to directly interpret the result a density 
distribution of photon emissions -- the normalization needs to be adapted. 
In extended events, the emission of the detected photons is not
restricted to a single point, but is still linked to the specific topology, providing 
a mixture of correlated and uncorrelated information. Therefore, the topological reconstruction 
performance can be enhanced by exploiting the present correlation.

To utilize this in the reconstruction, in the following we make the basic assumption that all 
detected photons originate from the same event topology, i.e., that their emission points are
restricted to a subvolume of the detector. By this we neglect spurious signals, e.g., from the 
photosensor dark noise or omnipresent radioactivity events, which contribute essentially the same 
way as scattered photons in the reconstruction. Doing so, we define a `probability mask' that defines
a certain prior knowledge of the event's energy deposition topology. It is expressed by a 
three-dimensional PDF $M(\mathbf{x})$ with the normalization condition
\begin{equation}
\int \limits_{V_{\mathrm{LS}}} M(\mathbf{x})  \mathrm{d} V \, \overset{!}{=} 1 \, .
\label{eq:ProbMaskNormCond}
\end{equation}
Like a `prior' in Bayesian statistics, the mask 
represents prior information on the spatial distribution of photon emissions. For example, it can be 
constructed based on outcomes of other, more coarse reconstruction methods. This helps to predefine 
a volume where to look for energy depositions before starting the computing-intensive topological 
reconstruction.

To apply the probability mask, which essentially reweights the spatial PDFs 
$\Phi_{j,k}(\mathbf{x})$ generated for every single photon prior to its normalization, 
$M(\mathbf{x})$ is included outside of the time integral in eqs.~(\ref{eq:SinglePhotPDF}) and 
(\ref{eq:NormFactor}), which then become
\begin{equation}
 \Phi_{j,k}(\mathbf{x}) = w_{j,k} \, M(\mathbf{x}) \,\varepsilon_j(\mathbf{x}) \int\limits_0^\infty
 \Phi_{t_s}(\Delta t) \,
 \Phi_{t_{\mathrm{ph}}}(t';\mathbf{x},\mathbf{r}_j) \, \mathrm{d} t'
 \label{eq:SinglePhotPDFwMask}
\end{equation}
and
\begin{equation}
  w_{j,k}^{-1} = \int\limits_{V_{\mathrm{LS}}}
  M(\mathbf{x})\,
  \varepsilon_j(\mathbf{x}) \int\limits_0^\infty
  \Phi_{t_s}(\Delta t) \,
  \Phi_{t_{\mathrm{ph}}}(t';\mathbf{x},\mathbf{r}_j) \, \mathrm{d} t' \mathrm{d} V \, .
  \label{eq:NormFactorwMask}
\end{equation}

In principle, the choice of the probability mask is free as long as it i) represents the prior 
knowledge on an event topology in an unbiased way and ii) is sufficiently smooth as a function of 
$\mathbf{x}$. The latter point is important to avoid reconstruction artifacts at sharp edges of 
the probability map due to over-enhancement of photon signals that just barely overlap with 
the mask's significant region. Typically, this happens for strongly 
delayed photons (e.g., due to scattering or slow scintillation) that carry no 
topological information on the event and thus are only noise to the reconstruction.

A natural way to create a probability mask for the topological reconstruction is the following: 
In a first step, the reconstruction is performed without a mask as described in section 
\ref{sec:Method.BasicIdea}. The outcome constitutes an unbiased representation of the event 
topology. This result can be converted into a probability mask $M(\mathbf{x})$ (additional 
smoothing can be applied) and is fed back into a second reconstruction run. A repetition of this 
procedure to improve the outcome thus makes the whole topological reconstruction an iterative 
method where $M(\mathbf{x})$ changes between iterations. 
However, it is essential to avoid the introduction of self-enhancement effects, i.e., a 
photosensor must not be biased by its own contribution to the probability mask created in previous 
iterations. In principle, one unique mask $M_j(\mathbf{x})$ for every photosensor has to be 
generated using information of all sensors but its own. However, to avoid this
computing-intensive procedure, we followed a more practical approach to mitigate self-enhancement effects,
which is described in paragraph iv of section \ref{sec:Analysis.RecoProcedure}.

\begin{figure}[ht!]
  \centering
  \begin{minipage}{0.98\textwidth}     
    \includegraphics[trim=0.1cm 0.1cm 0.0cm 0.1cm,clip=true,width=\textwidth]
      {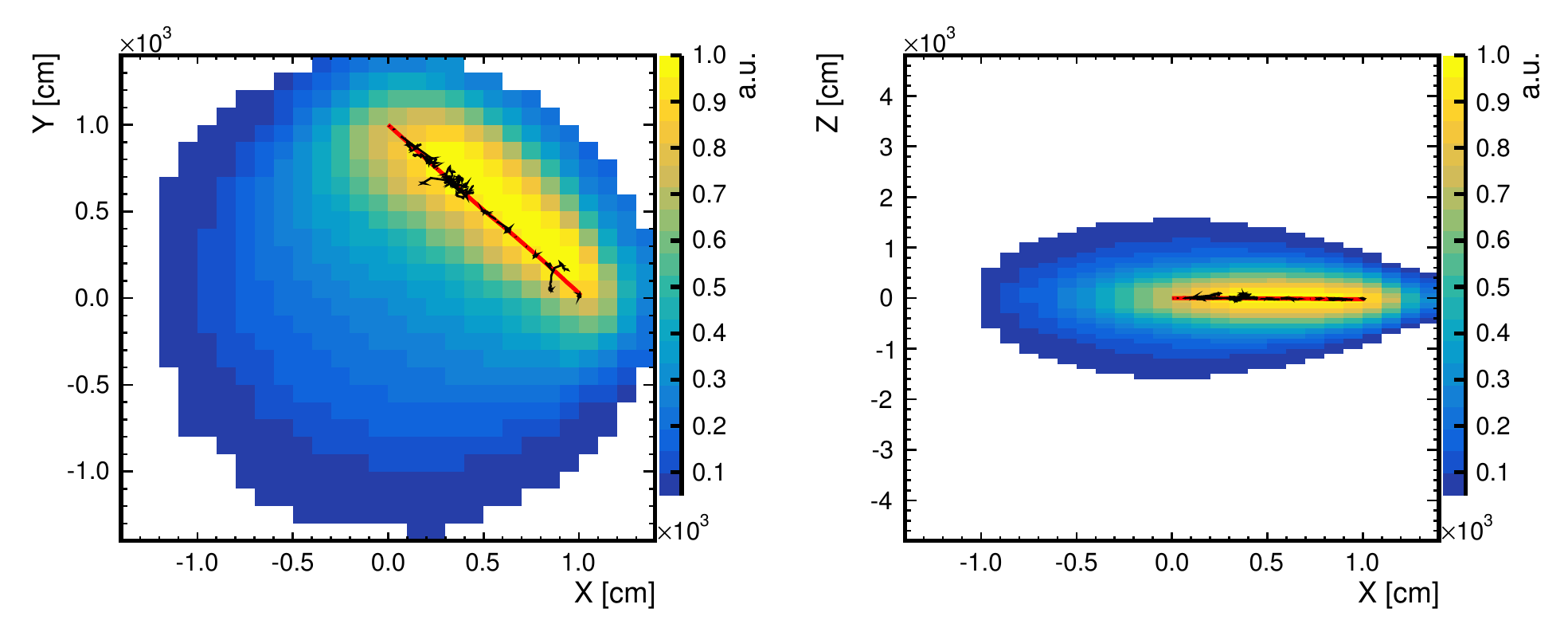}
  \end{minipage}
  ~
  \begin{minipage}{0.98\textwidth}     
    \includegraphics[trim=0.1cm 0.1cm 0.0cm 0.1cm,clip=true,width=\textwidth]
      {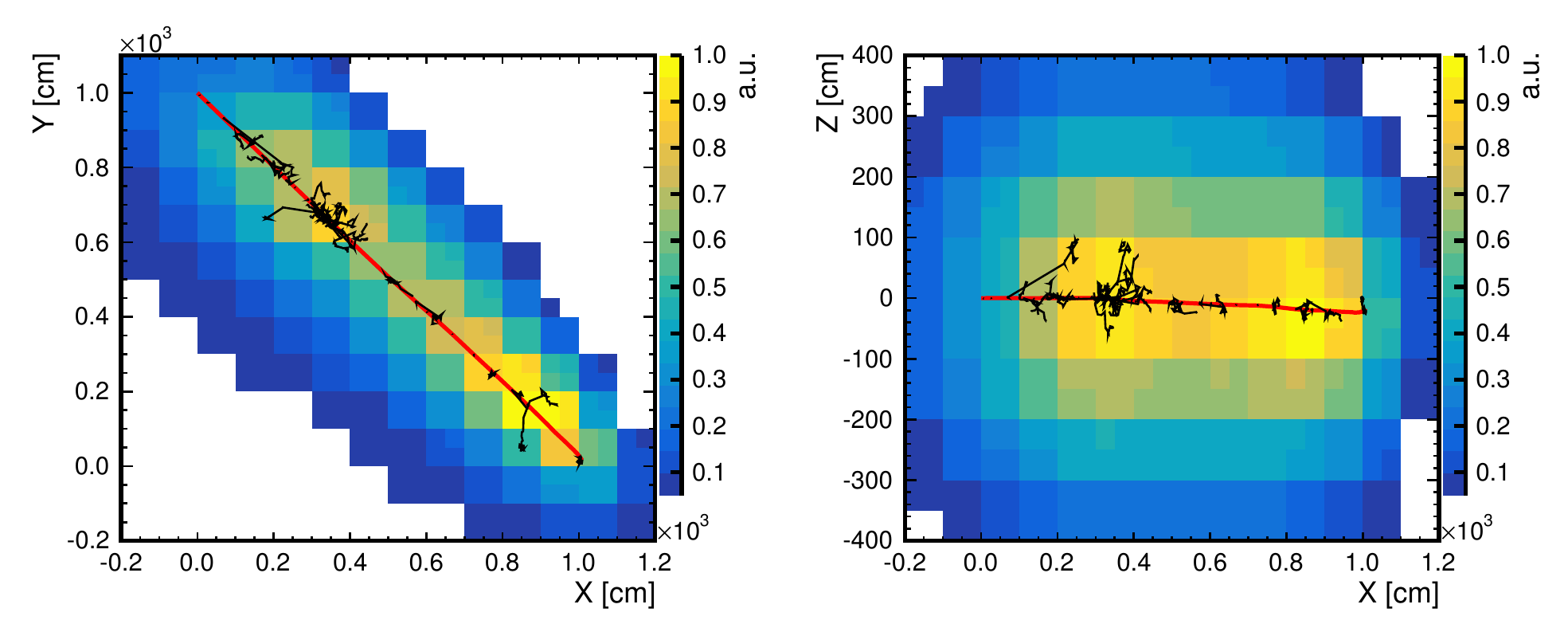}
  \end{minipage}
  ~
  \begin{minipage}{0.98\textwidth}
    \includegraphics[trim=0.1cm 0.1cm 0.0cm 0.1cm,clip=true,width=\textwidth]
      {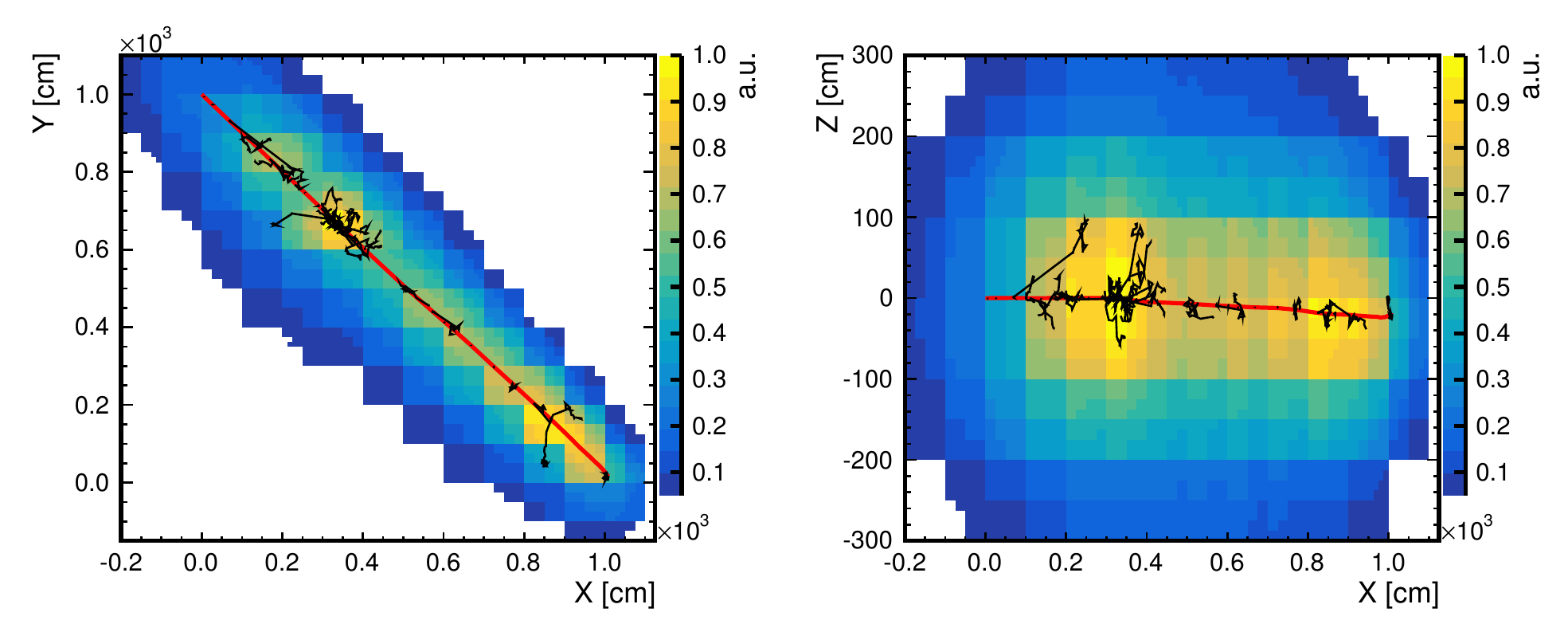}
  \end{minipage}
  \caption{
Reconstruction results after the iterations 0 (top), 8 (middle) and 21 (bottom) for a 
simulated muon with 3\,GeV initial kinetic energy in the cylindrical LENA detector projected 
along the symmetry axis (left) or a radial $y$-axis (right). The primary particle started at 
$(0,\,1000,\,0)$\,cm in the direction $(1,\, -1,\, 0)$. Both the projected tracks of the 
primary particle (red) and of secondary particles (black) are shown. Note that both the axis scales 
and the sizes of the cells change due to the selection of a region of interest and the refinement 
of the reconstruction mesh. Moreover, the cell content is given in a.u. and rescaled such that the 
maximum content is 1. Some details on the actual reconstruction procedure are given in section 
\ref{sec:Analysis.RecoProcedure}.}
  \label{fig:RecoItrResults}
\end{figure}

The power of this iterative process is illustrated in figure \ref{fig:RecoItrResults}. Signal photons 
are clearly associated to a subvolume of the detector defined by the event topology. What is more, the
result reflects the local differential energy loss $\mathrm{d}E/\mathrm{d}x$, the basis to
facilitate more localized spatial vetoes around showers for cosmogenic background rejection.

\section{Performance analysis}
\label{sec:Analysis}

We tested the performance of the topological reconstruction with a Monte Carlo (MC) data set prepared
with the LENA detector simulation described in section \ref{sec:Analysis.DetSim}. The sample of 
simulated single muon events is the topic of section \ref{sec:Analysis.EventSample}.

A key feature of the iterative topological reconstruction method is its flexibility 
concerning the actual implementation as an algorithm. Changeable settings include the reference point / 
time, the number of iterations and especially the construction of a probability mask prior to 
a new iteration. They make the method highly customizable. The concrete implementation applied here is 
described in section \ref{sec:Analysis.RecoProcedure}. The quantitative analysis of the three-dimensional 
data in the final reconstruction outcome, 
$\Gamma_{\mathrm{em}}(\mathbf{x})$, that extracts descriptive physics parameters (e.g., angular 
and spatial resolutions), is the topic of section \ref{sec:Analysis.AnaProcedure}. Details can also be 
found in~\cite{Lorenz:2016_PhDThesis}. Results are presented in section \ref{sec:Analysis.Results}.

\subsection{Detector simulation}
\label{sec:Analysis.DetSim} 

The performance study is based on a sample of single muon events produced with the help of the
GEANT4-based\footnote{Used version: 4.9.6 Patch 02} MC simulation of the LENA~\cite{Wurm:2011_LENA} 
detector. Details about the simulation, like the optical model of the LS, 
can be found in~\cite{Moellenberg2013_PhDThesis}. A comprehensive description of the implemented 
detector geometry is given in~\cite{Lorenz:2016_PhDThesis}. The following only summarizes points 
relevant for this analysis.

In the LENA detector design, a cylindrical volume of 96\,m height and 14\,m radius containing about
50\,kt of LAB-based LS~\cite{Helm:2011_LAB} serves as designated neutrino target. The decay time 
distribution of the LS is modeled according to eq.~(\ref{eq:ScintEmissionTimingPDF}) 
with three fluorescence components, where the fastest, dominant component for electron excitation
has $\tau=4.6$\,ns ($\omega = 71\%$) and the slowest component has 156\,ns (7\%)~\cite{Marrodan:2008_PhDThesis}.

To save computing time, the optical model describing photon propagation 
in LAB neglects wavelength-dependent effects: All optical quantities, like the refractive index and individual 
scattering and absorption lengths, are effective values regarding the primary wavelength of
the scintillation light transport at 430\,nm. The model implements absorption (20\,m absorption length)
and anisotropic Rayleigh scattering from LAB molecules, reflected by the pair of isotropic (60\,m) and 
anisotropic (20\,m) scattering lengths~\cite{Wurm:2010_OptScat}. To simplify 
the situation for this study, we disabled light production via the Cherenkov effect as neither the 
considered LS mixture (transparency) nor the photosensor system (wavelength sensitivity, time 
resolution) in the baseline setup of LENA were optimized for Cherenkov photon detection / 
discrimination. The impacts of the promptly emitted and directional
Cherenkov cone for reconstruction in suitable hybrid detector setups are under investigation.

In LENA, scintillation light is detected with 30\! 542 PMTs of 12\,inch diameter and attached LCs.
To reduce computing time, the LCs are not modeled geometrically. Instead, 
flat photosensitive disks with an area corresponding to the aperture of an LC and angle-dependent
detection efficiency are introduced. The chosen LCs reproduce an acceptance similar to the Borexino-type LC 
with $\sim 44^{\circ}$ critical incident angle~\cite{Oberauer:2003_BxLightGuide}. To further 
reduce computing time, we assign 100\% quantum 
efficiency (instead of the baseline 20\%) to the PMTs and in exchange reduce the scintillation light 
yield to about 2\! 000\,$\mathrm{MeV}^{-1}$ (from about 10\! 000\,$\mathrm{MeV}^{-1}$) to arrive at 
the same of 265
photo-electrons per MeV for the center of the detector. All photons detected by the PMTs are saved 
to the output file. To include a basic reproduction of the PMT transit time spread, the obtained 
photon arrival times were smeared according to a Gaussian PDF with 1\,ns standard deviation. No 
further PMT or electronics effects, like the effects of a FADC-waveform, were considered, 
neglecting photon counting inefficiencies and limitations in the dynamic range of the 
read-out.

\subsection{Event sample}
\label{sec:Analysis.EventSample} 

\begin{figure}[b!t]
  \centering
  \begin{minipage}{0.48\textwidth}
    \includegraphics[trim=0.1cm 0.1cm 0.0cm 0.1cm,clip=true,width=\textwidth]{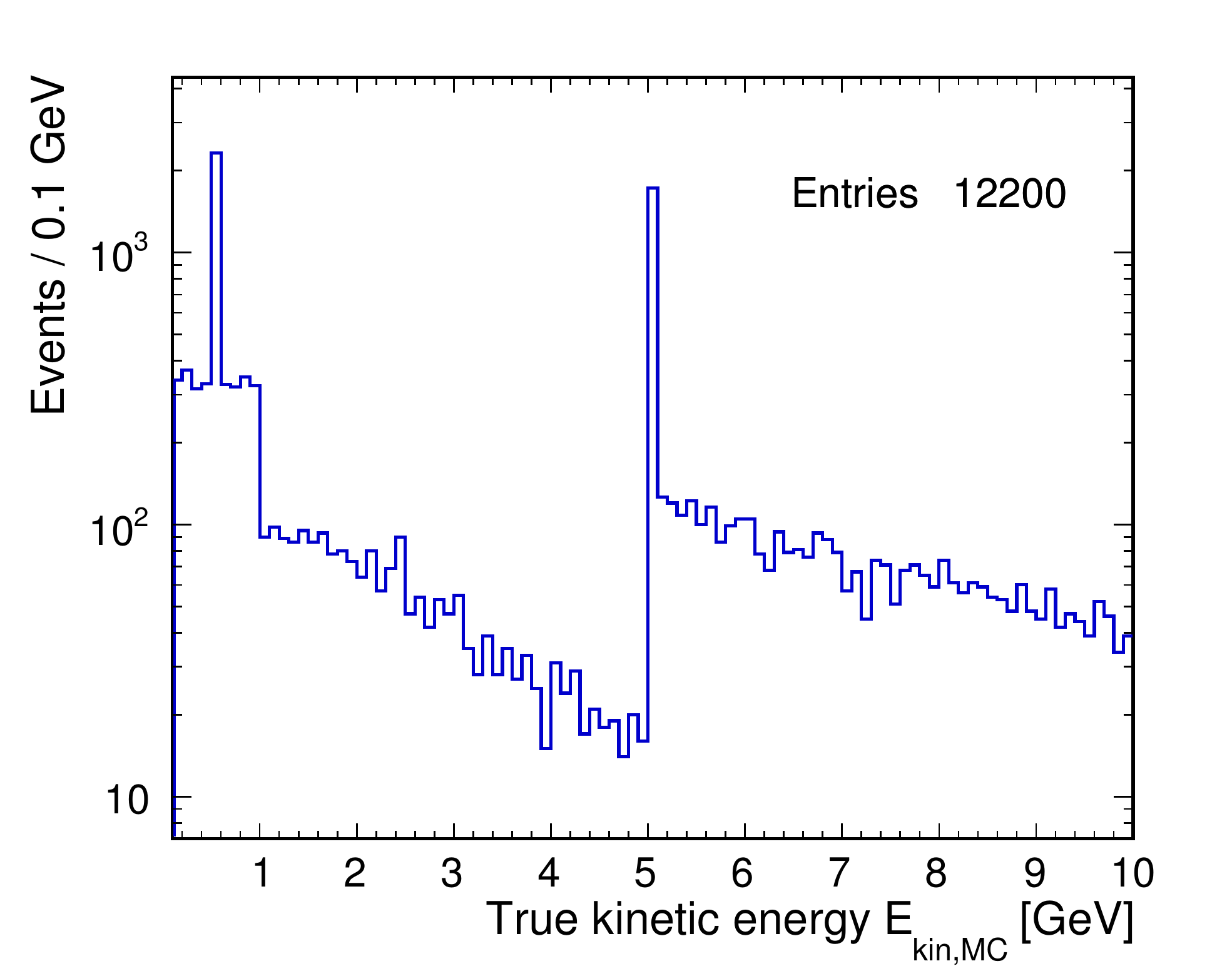}
  \end{minipage}
  ~
  \begin{minipage}{0.48\textwidth}
    \includegraphics[trim=0.1cm 0.1cm 0.0cm 0.1cm,clip=true,width=\textwidth]{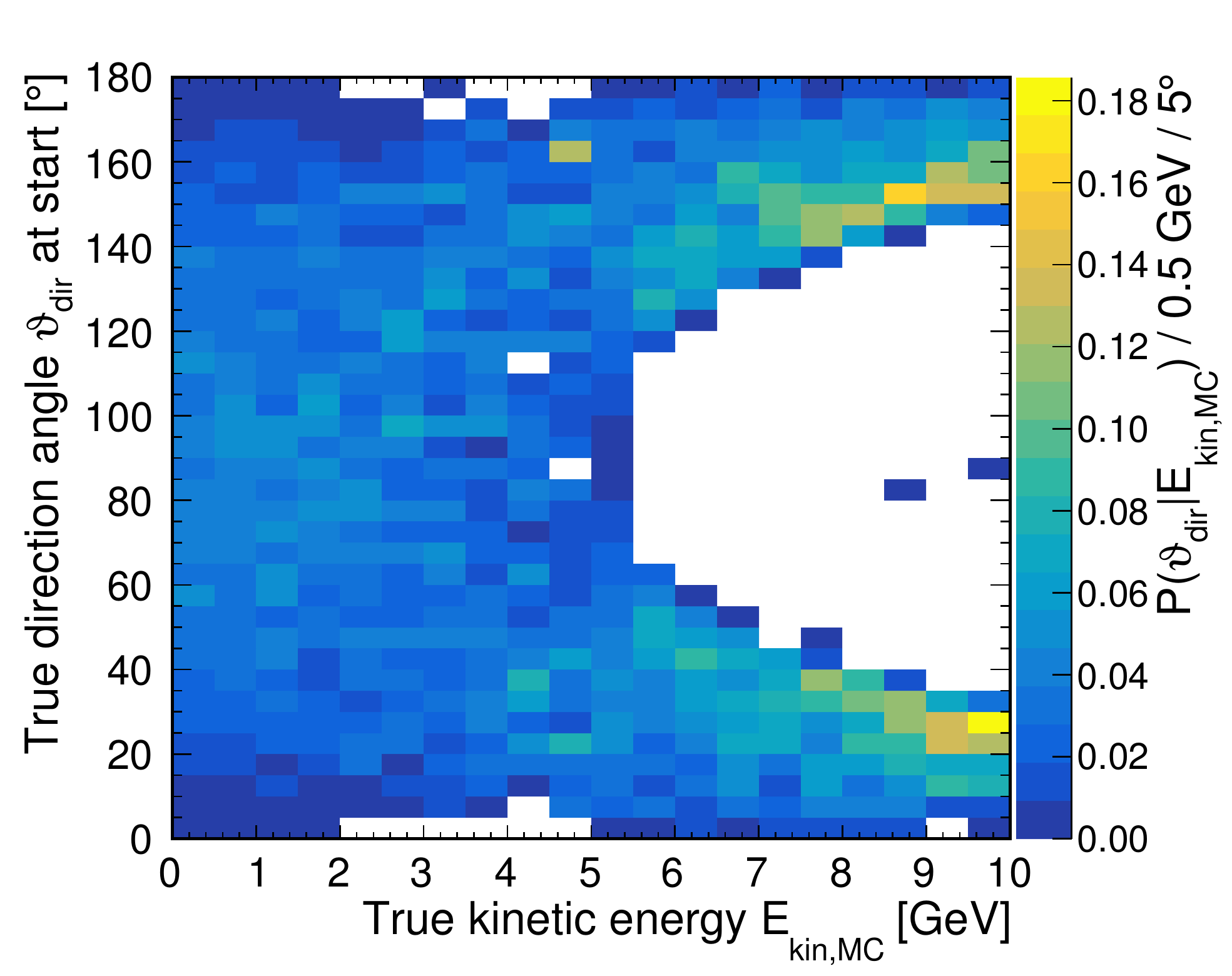}
  \end{minipage}
    \caption{
Left: Distribution of the initial kinetic energy $E_{\mathrm{kin,MC}}$ of the primary muons in the final 
event sample used to evaluate the performance of the reconstruction method 
presented in section \ref{sec:Method}. The total sample with 12\! 200 events is composed of five 
sub-samples with different energy ranges: [0.1,\,1]\,GeV (3\! 000), [1,\,5]\,GeV (2\! 
000), [5,\,10]\,GeV (3\! 580), 0.5\,GeV (2\! 000) and 5\,GeV (1\! 620).
Right: Distribution of the polar angle $\vartheta_{\mathrm{dir}}$ of the initial muon direction versus the 
particle's initial kinetic energy $E_{\mathrm{kin,MC}}$. The polar angle is relative to the 
cylinder axis of the LENA detector (upward direction: $\vartheta_{\mathrm{dir}} =0^{\circ}$). The 
binned distributions of $\vartheta_{\mathrm{dir}}$ were renormalized to one per 
$E_{\mathrm{kin,MC}}$ bin.}
  \label{fig:TrueDists}
\end{figure}

The LENA detector simulation was used to generate a sample of muon 
events for the performance test of the reconstruction. It is divided into five sub-samples: Three 
sub-samples individually cover the initial kinetic energy ranges 0.1 to 1\,GeV, 1 to 5\,GeV and 
5 to 10\,GeV. The energy value of the primary particle was drawn from a flat distribution in each 
range. Two sub-samples with higher statistics contain muons at fixed initial kinetic energies, 
either 0.5 or 5\,GeV. Each primary muon was simulated with a random momentum direction. 
To limit the study to fully-contained events and to avoid unnecessary complications 
due to end-cap effects, we afterwards selected events based on MC truth information for the analysis.
The selection required that both start and end point of a muon track are confined to a cylindrical volume 
of 50 m height and 24\,m diameter centered on the target volume of LENA. As a consequence of this event selection, 
longer tracks from muons of higher 
energy are more likely to be excluded and the track direction of remaining higher energy muons is more
aligned with the symmetry axis of the LENA detector. Both aspects are reflected in figure \ref{fig:TrueDists}:
The left side shows the true kinetic energy distribution of the final muon sample, which has the non-flat regions
1--5\,GeV and 5--10\,GeV due to the implicit cutting on energy. The right side shows distributions of the polar angle $\vartheta_{\mathrm{dir}}$ 
of the initial muon direction per true kinetic energy bin, which become depopulated in an increasingly broad
angle range around the horizontal direction $\vartheta_{\mathrm{dir}} =9 0^{\circ}$ at higher energies due to the 
more and more vertical alignment of the long muon tracks.

Note that while we constrained the analysis for simplicity to fully-contained muon events, the 
presented reconstruction method is not limited to this event class.

\subsection{Reconstruction procedure}
\label{sec:Analysis.RecoProcedure}

The description of the reconstruction procedure evolves around the following aspects: 
i) the required reference parameters 
$\mathbf{r}_{\mathrm{ref}}$ and $t_{\mathrm{ref}}$, ii) the computation of the single photon PDFs
$\Phi_{j,k}(\mathbf{x})$ in eq.~(\ref{eq:SinglePhotPDF}), iii) the computations on a discrete 
domain, i.e., a mesh, iv) the generation of the probability masks $M_j(\mathbf{x})$ and v) the fixed 
sequence of reconstruction iterations with their individual settings.

\paragraph{i) Reference parameters} Both the reference point $\mathbf{r}_{\mathrm{ref}}$ 
and the reference time $t_{\mathrm{ref}}$ in eq.~(\ref{eq:RecoTimeModel}) were obtained by adding 
random shifts $\Delta \mathbf{r}$  and $\Delta t$ to the true MC parameters at the muon start 
point. The spatial shift was composed of three individual shifts, $\Delta \mathbf{r} = (\Delta x, 
\Delta y, \Delta z)$, each drawn from a normal distribution $\mathcal{N}(0,10$ cm$)$ \cite{Wurm:2011_LENA}. 
This yields a mean shift $\overline{|\Delta \mathbf{r}|}$ of the reference point with respect to the 
true start point of about 16\,cm. Similarly, the temporal shift $\Delta t$ was drawn from a 
normal distribution $\mathcal{N}(0,1$\,ns$)$. For comparison, a start point 
resolution $<7$ cm (note the resolution definition) was found in~\cite{Hellgartner:2015_PhDThesis} 
with a likelihood-based track fit applied to muons with kinetic energies
$\leq 1$\,GeV. The start time resolution was found to be $<0.14$\,ns. Therefore, our assumptions are conservative. 
The choice of the reference parameters close to the
start point simplifies the description. It is not a necessary condition as the topological reconstruction 
can be extended both ``forward'' and ``backward in time'' (cf. eq.~(\ref{eq:RecoTimeModel})). During
the iteration, intermediate reconstruction outcomes for $\Gamma_{\mathrm{em}}(\mathbf{x})$ can be used to 
extract new reference parameters corresponding to the start point of a track. For realistic detector
settings, e.g., in the reconstruction of cosmic muons, the reference parameters can be provided by external 
(veto) detectors or other reconstruction methods, for example, the 
``backtracking-algorithm''~\cite{Hellgartner:2015_PhDThesis}.

\paragraph{ii) Single photon PDFs} The computation of the single photon PDFs
$\Phi_{j,k}(\mathbf{x})$ as defined in eq.~(\ref{eq:SinglePhotPDFwMask}) is time-consuming due to the 
integration over different possible photon travel times, which are weighted by the PDF 
$\Phi_{t_{\mathrm{ph}}}(t;\mathbf{x},\mathbf{r}_j)$ including the optical model. However, 
$\Phi_{t_{\mathrm{ph}}}(t;\mathbf{x},\mathbf{r}_j)$ can be precomputed and stored in 
three-dimensional look-up tables (LUTs) as a function of time and the position relative to the photosensor's 
reference point. Depending on the level of detail (photon scattering!) included in the 
tabulated data and the geometry of the detector, the number of required LUTs ranges from one per 
single photosensor to one per photosensor type. For our MC study, we made two simplifications with 
respect to this LUT: First, we did not tabulate the entire PDF
$\Phi_{t_{\mathrm{ph}}}(t;\mathbf{x},\mathbf{r}_j)$ but only the mean arrival 
time $\overline{t_{\mathrm{ph}}}$ of a photon from a given position relative to the sensor. 
This removed the integration from eq.~(\ref{eq:SinglePhotPDFwMask}) and reduced the LUT to two dimensions.
Second, we treated all scattered photons as absorbed and thus not detected, 
making the LUT independent of the actual photosensor, i.e, only one LUT for all sensors was needed. 
This was based on the assumption that scattered photons essentially do not contribute meaningful information to
the reconstruction due to their delayed arrival and random scattering direction but merely smear out 
the topology of an event.

Similar to the mean photon arrival time, the spatial detection efficiency
$\varepsilon_j(\mathbf{x})$ in eq.~(\ref{eq:SinglePhotPDFwMask}) can be precomputed and tabulated as
well in two-dimensional LUTs as a function of the position relative to the photosensor.
By neglecting scattered photons in the computation of $\varepsilon_j(\mathbf{x})$, only a single LUT
becomes sufficient to describe all PMTs. However, this leads to underestimations of the
single detection efficiencies $\varepsilon_j(\mathbf{x})$ and
finally of the global detection efficiency $\varepsilon(\mathbf{x})$ in eq.~(\ref{eq:LocalDefEff}).

\begin{figure}[b!t]
  \centering
    \includegraphics[width=0.65\textwidth]{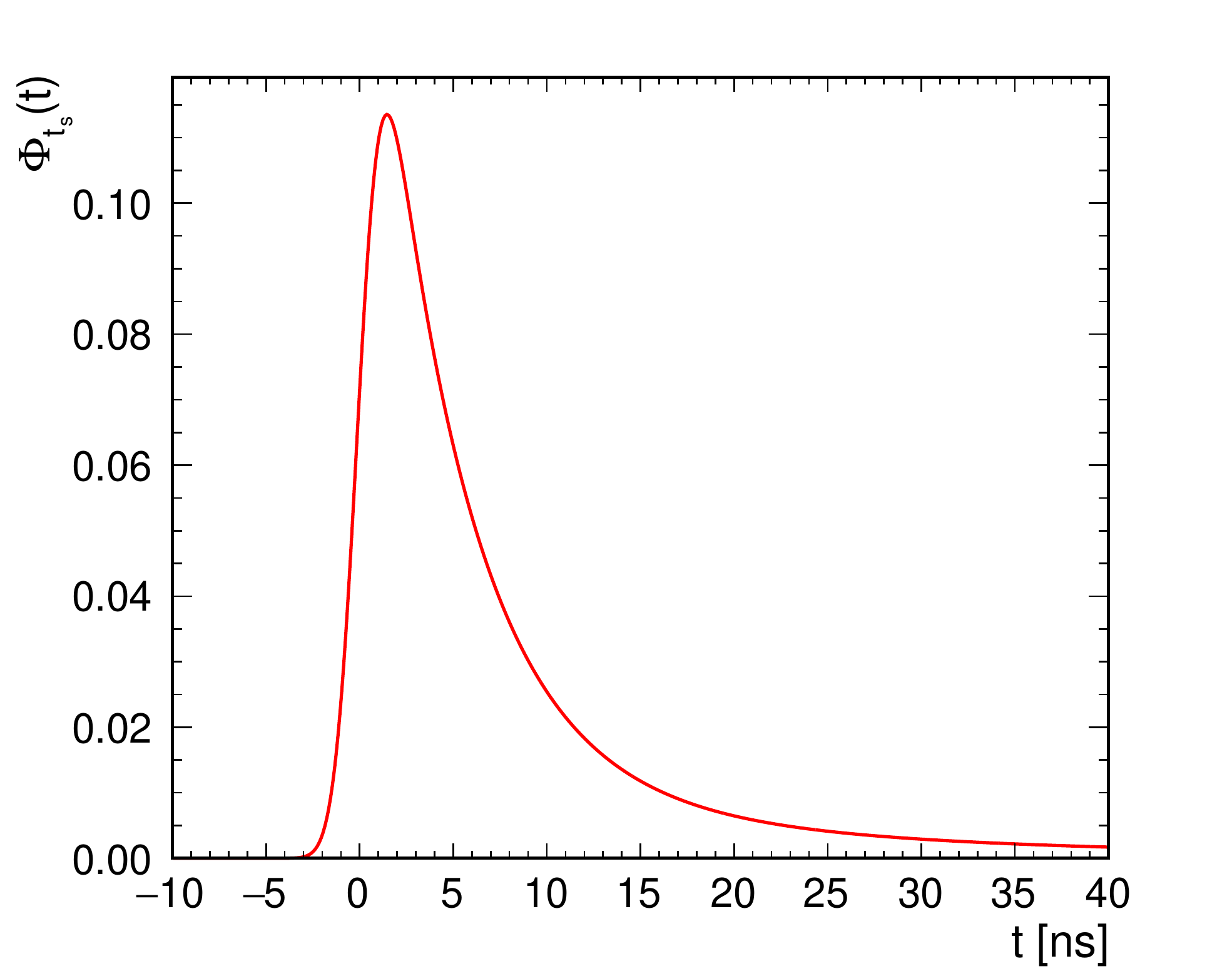}
    \caption{
The PDF $\Phi_{t_s}(t)$ as we used to model the random time contributions $t_s$ in 
eq.~(\ref{eq:RecoTimeModel}). It is the convolution of a normal function 
$\mathcal{N}(0,1\,\mathrm{ns})$ to account for the photosensor time jitter and the PDF
$\Phi_{\textrm{em}}(t)$ in eq.~(\ref{eq:ScintEmissionTimingPDF}) to model the scintillator timing. 
The $\tau_i$ and $\omega_i$ are (4.6\,ns, 18.0\,ns, 156\,ns) and (0.71, 0.22, 0.07), respectively.}
  \label{fig:SmearPDF}
\end{figure}

For the PDF $\Phi_{t_s}(t)$ in eq.~(\ref{eq:SinglePhotPDFwMask}), which describes random time 
contributions $t_s$ to a photon's expected signal time $\hat{t}_j(\mathbf{x})$ from the photosensor 
time jitter and the scintillator decay, we used the model shown in figure \ref{fig:SmearPDF}: It is 
a convolution of a normal function to account for the photosensor time jitter and the PDF 
$\Phi_{\textrm{em}}(t)$ in eq.~(\ref{eq:ScintEmissionTimingPDF}) to reflect scintillator timing.
Due to the simplifications used in the consideration of different photon travel times, 
$\Phi_{j,k}(\mathbf{x})$ reduces to
\begin{equation}
 \Phi_{j,k}(\mathbf{x}) = w_{j,k} \, M(\mathbf{x}) \, \varepsilon_j(\mathbf{x}) \, \Phi_{t_s}(\Delta t)\, ,
 \label{eq:OurRecoFunction}
\end{equation}
with the difference between measured and expected signal time being $\Delta t = t_{j,k} - 
\hat{t}_j(\mathbf{x}; t_{\mathrm{ph}} = \overline{t_{\mathrm{ph}}}, t_s = 0)$. The calculation of 
the normalization factor $w_{j,k}$ in eq.~(\ref{eq:NormFactorwMask}) was adjusted accordingly.

\paragraph{iii) Reconstruction mesh} Since the calculations described in section \ref{sec:Method} 
are computing intensive, the reconstruction was not performed continuously in $\mathbf{x}$ but 
on a discrete, three-dimensional mesh covering the LS volume. The 
mesh itself was composed of cubic cells and had the minimum extent to fully enclose the 
cylindrical active volume of LENA.
All calculations were performed with respect to the center points of the 
cells. During the reconstruction, the cell size of the initially coarse mesh was decreased over
subsequent iterations to arrive finally at a spatial resolution relevant for the reconstruction 
result. To save computing time, computations were focused on the region of interest by reducing
the cell size only in the region of the muon track. Moreover, they were limited to cells fully or
partially contained in the LS volume. For the latter, the algorithm averaged over smaller, virtual
sub-cells (cells without space in memory) of the real cell that were completely inside the LS volume. 
Details on the implementation of the adaptive mesh refinement and the computations with virtual 
cells are given in~\cite{Lorenz:2016_PhDThesis}. 

\paragraph{iv) Probability masks} In general, the outcome $\Gamma_{\mathrm{em}}(\mathbf{x})$ of an 
iteration served as probability mask for the next computations. The 0th step used no probability mask. 
Our approach to avoid the self-enhancement mentioned in section \ref{sec:Method.Iteration} is 
described in the following paragraph.

\paragraph{v) Iteration sequence} We used a total of $N_i=22$ iterations over which the grid constant
of the mesh with cubic cells was stepwise reduced in the refined regions. We performed eight 
iterations with 100\,cm, eight iterations with 50\,cm, five iterations with 25\,cm and one iteration 
with 12.5\,cm. In order to avoid self-enhancement, the computations according to 
eq.~(\ref{eq:OurRecoFunction}) in subsequent iterations were based on alternating sub-sets of hit 
photosensors.\footnote{Unhit photosensors can be used to shape a probability mask
by evaluating an intermediate iteration result $\Gamma_{\mathrm{em}}(\mathbf{x})$ from all hit sensors: For a given
combination of photosensor and mesh cell with center $\mathbf{x}$, one can calculate the probability that the 
reconstructed density value $\Gamma_{\mathrm{em}}(\mathbf{x})$ created no photon hit at the unhit sensor.
By multiplying each cell value with the corresponding product of probabilities from all unhit photosensors,
one can finally create a smoothened probability mask. This method is useful to remove coincidentally 
occurring artifacts in $\Gamma_{\mathrm{em}}(\mathbf{x})$. However, it should solely be used to create
probability masks for further iterations. Its application to the final result would destroy information by
unnecessarily diminishing the result's contrast.} 
Only the last two iterations utilized the information from all hit photosensors.
In any case, all photosensors were taken into account for the computation of the global detection 
efficiency according to eq.~(\ref{eq:LocalDefEff}). Although this approach does not prevent 
self-enhancement entirely, it significantly mitigates its impact. Further improvement can be expected 
if one increases the number of photosensor sub-sets.

The iteration sequence can be summarized as follows:
Let $0 \leq i \leq N_i-1$ be the index of the current iteration.
\begin{enumerate}
 \item Iteration with even-indexed photosensors; increment $i$ by 1.\\
       (If $i=0$, use no probability mask.)
 \item Create probability mask from intermediate result.
 \item Iteration with odd-indexed photosensors; increment $i$ by 1.
 \item Refine mesh in region of interest if cell size changes for next iteration.
 \item Create probability mask from intermediate result.
 \item Repeat the previous steps as long as $i < N_i-2$.
 \item Iteration with all photosensors; increment $i$ by 1.
 \item Refine mesh in region of interest if cell size changes for next iteration.
 \item Create probability mask from intermediate result.
 \item Final iteration with all photosensors; increment $i$ by 1.
\end{enumerate}
The flexibility to tune the iterative reconstruction of an event (number of iterations, 
refinement steps for cell size, manipulation of the probability mask between iterations etc.)
gives significant power to the reconstruction technique and makes it highly versatile. As a consequence,
there is nothing like a \textit{standard topological reconstruction procedure}, making the iteration sequence 
presented above just one (not yet optimized) implementation.

\subsection{Analysis procedure}
\label{sec:Analysis.AnaProcedure}

For the interpretation of the reconstruction result $\Gamma_{\mathrm{em}}(\mathbf{x})$, it is 
crucial to extract descriptive parameters for the events. In our case of fully contained muons, 
we decided to select the start point $\mathbf{x}_s$, the track direction $\mathbf{d}(\varphi,\theta)$,
its end point $\mathbf{x}_e$ and an estimator for the total number of emitted photons $N_{em}$. 
With a view on cosmogenic background rejection based on spatial vetoes, one critical issue
is the accurate localization of muon tracks in the LS volume. Especially a precise determination of a
muon track's end point is important to construct sufficiently long veto regions for stopping muons, i.e., 
muons that enter the detector from outside but stop inside the LS volume. These aspects are evaluated in 
our study in terms of resolutions for start and end point of fully-contained muon tracks. While
the above parameter choices provide adequate description 
for a minimum-ionizing muon track, in principle a broad range of tools from image and 
three-dimensional data analysis can be applied to $\Gamma_{\mathrm{em}}(\mathbf{x})$ to extract 
higher-level information ($\mathrm{d}E/\mathrm{d}x$, shower formation etc.).

In a first step, the event region in the reconstruction result has to be determined, i.e.,
the collection of cells in $\Gamma_{\mathrm{em}}(\mathbf{x})$ that are considered part of the 
event topology (the muon track). This is a non-trivial task because the contrast between 
important and unimportant regions in $\Gamma_{\mathrm{em}}(\mathbf{x})$ depends on multiple 
factors (total energy deposition, goodness of reference parameters etc.) and requires dedicated 
studies and tuning. Here, we define the event region based on a relative threshold: All 
cells with a value greater than 2\% of the maximum cell value were considered to be part of 
the muon event. Note that by increasing the threshold value prominent regions in 
$\Gamma_{\mathrm{em}}(\mathbf{x})$ with increased energy deposition can be selected (e.g., 
bottom left of figure \ref{fig:RecoItrResults}). Such a selection would provide enhanced sensitivity
to cosmogenic radioisotope production in showering muon events and allows to develop more focused 
spatial vetoes around the showers.

To reconstruct the mean direction of the track, we fitted a line defined by a point of origin $\mathbf{s}$ 
and the direction angles $\varphi$ and $\theta$ in spherical coordinates through the selected event 
region. This was done by minimizing
\begin{equation}
 f(\mathbf{s},\varphi,\theta) = \ln \sum\limits_i \left( \Gamma_{\mathrm{em}}(\mathbf{x}_i) \,
 \cdot D_i (\mathbf{x}_i;\mathbf{s},\varphi,\theta) \right)^2
\end{equation}
over all cells in the event region. Here, $ D_i(\mathbf{x}_i;\mathbf{s},\varphi,\theta)$ is the 
orthogonal distance between the line hypothesis and the current cell with center point 
$\mathbf{x}_i$, which is weighted by the cell content 
$\Gamma_{\mathrm{em}}(\mathbf{x}_i)$. Using the outcome of the three-dimensional line fit, we 
reconstructed $\mathbf{x}_s$ as the line point having the smallest distance to 
the reference point $\mathbf{x}_{\mathrm{ref}}$ for the reconstruction, relying on the previous 
assumption that $\mathbf{x}_{\mathrm{ref}}$ is close to $\mathbf{x}_s$. For $\mathbf{x}_e$ we took 
the point where the fitted line exits the event region at the furthest distance to $\mathbf{x}_{\mathrm{ref}}$. 
Finally, $N_{em}$ was estimated by the volume integral of $\Gamma_{\mathrm{em}}$ in the event region. 
The result was also used to determine the energy resolution.

\subsection{Results}
\label{sec:Analysis.Results}

Results for the muon track reconstruction are presented in terms of the resolution for the track
direction $\mathbf{d}(\varphi,\theta)$, start ($\mathbf{x}_s$) and end point ($\mathbf{x}_e$) as well
as the reconstructed number of photon emissions ($N_{em}$) / event energy. In the following, 
reconstructed quantities are marked with a hat.

\begin{figure}[b!t]
  \centering
  \begin{minipage}{0.495\textwidth}     
    \includegraphics[trim=0.3cm 0.0cm 0.7cm 0.7cm, clip=true, width=\textwidth]
    {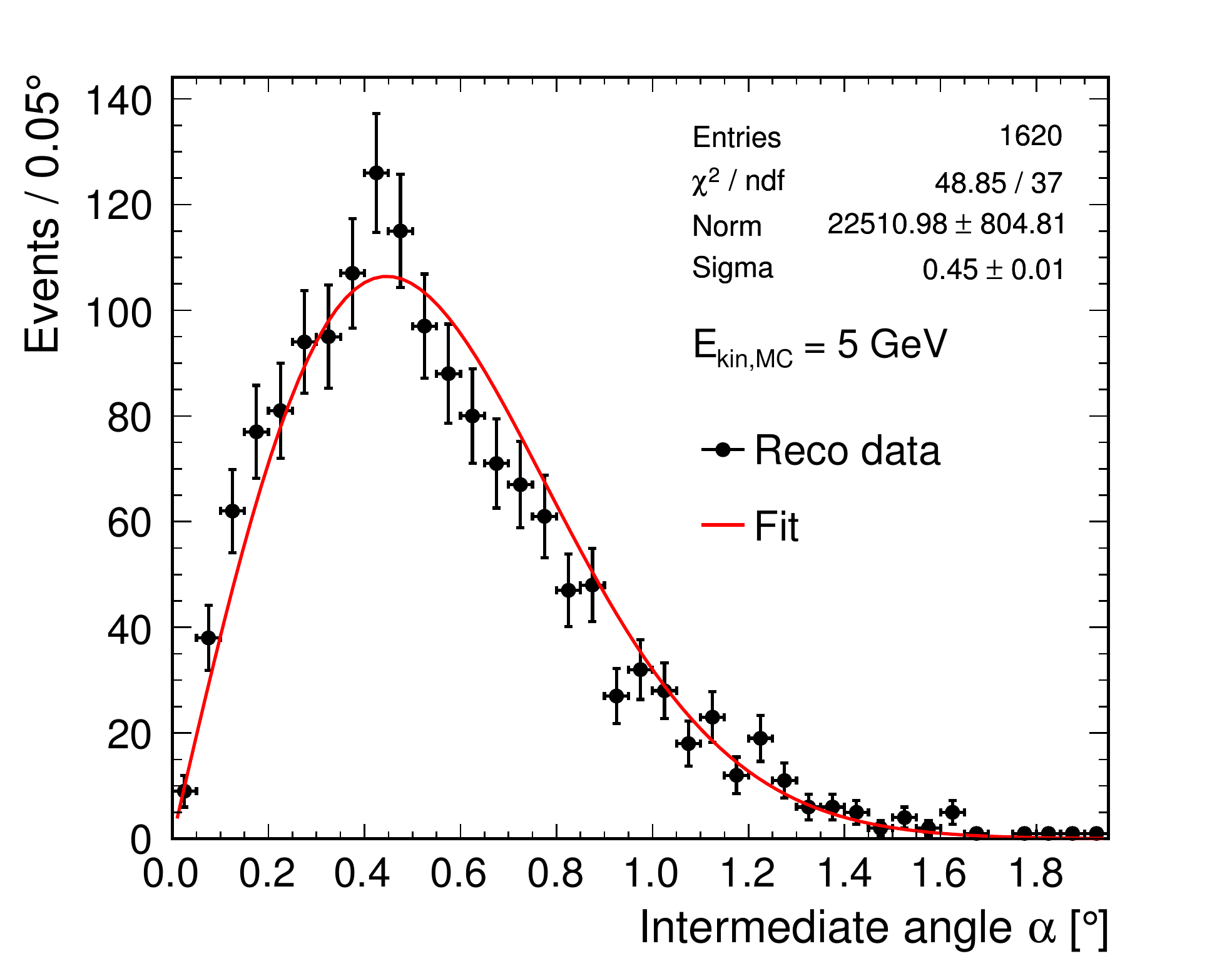}
  \end{minipage}
  \begin{minipage}{0.495\textwidth}
    \includegraphics[trim=0.3cm 0.0cm 0.7cm 0.7cm, clip=true, width=\textwidth]
    {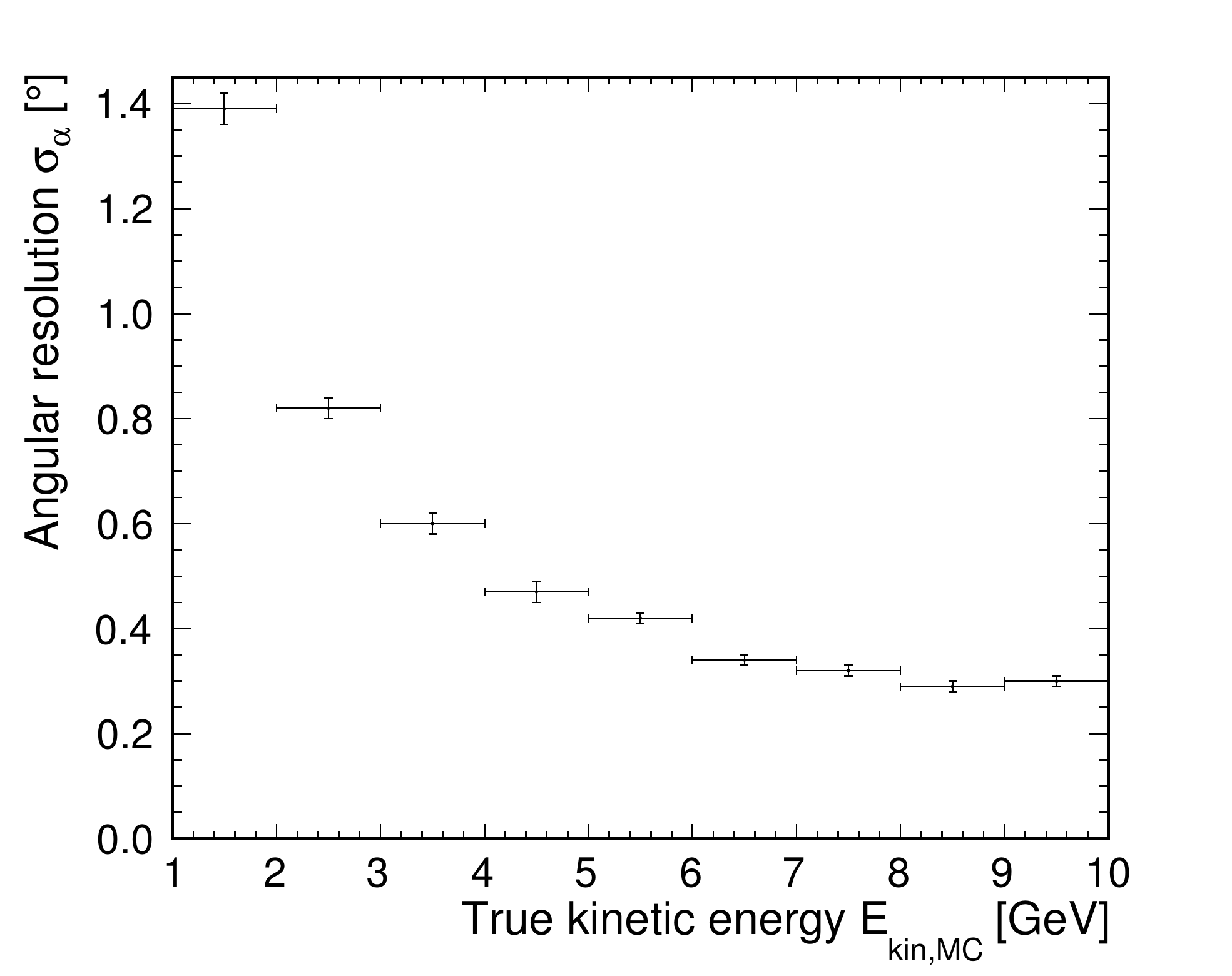}
  \end{minipage}
  \caption{
Left: Distribution of the intermediate angle $\alpha$ between reconstructed and true muon track for 
the high-statistics muon sample with $E_{\mathrm{kin,MC}} = 5$\,GeV true kinetic energy.
Vertical error bars only indicate statistical errors. A binned likelihood fit to the
distribution with eq.~(\ref{eq:AngResFitFcn}) was made. At larger angles, there are 52
outliers. Right: Angular resolution $\sigma_{\alpha}$ as a function of the true kinetic energy
$E_{\mathrm{kin,MC}}$ in the range from 1 to 10\,GeV. The resolutions were determined by binned 
likelihood fits of eq.~(\ref{eq:AngResFitFcn}) to distributions like the one shown on the left 
side. Vertical error bars indicate errors on $\sigma_{\alpha}$ returned from the fits.}
  \label{fig:ResultInterAngle}
\end{figure}

\paragraph{Angular resolution} To assess the angular resolution, we compared the reconstructed 
direction $\hat{\mathbf{d}}= \hat{\mathbf{x}}_e - \hat{\mathbf{x}}_s / | \hat{\mathbf{x}}_e - 
\hat{\mathbf{x}}_s |$  to the muon track direction 
$\mathbf{d}_{\mathrm{MC}} = \mathbf{x}_e - \mathbf{x}_s / |\mathbf{x}_e - \mathbf{x}_s |$ 
from MC. The intermediate angle is calculated as 
\begin{equation}
 \alpha = \arccos \left( \frac{\hat{\mathbf{d}} \cdot \mathbf{d}_{\mathrm{MC}}}
          {|\hat{\mathbf{d}}| \, |\mathbf{d}_{\mathrm{MC}} |} \right) \, .
\end{equation}
For a given energy range, we fitted the resulting distributions of $\alpha$ with the function
\begin{equation}
 F(\alpha) = \sin(\alpha) \, A \, \exp\left( - \frac{\alpha^2}{2\sigma^2_{\alpha}}\right) \, ,
\label{eq:AngResFitFcn}
\end{equation}
where we defined the fit parameter $\sigma_{\alpha}$ as the angular resolution. The factor
$\sin(\alpha)$ accounts for the solid angle dependence and the exponential function with 
normalization $A$ describes the resolution function, i.e., a Gaussian function around zero.
This description yields good results in the energy range above 1\,GeV (for below 1\,GeV 
see remarks at the end of this section):
The left side of figure~\ref{fig:ResultInterAngle} shows the distribution of 
$\alpha$ for the high-statistics sample at $E_{\mathrm{kin,MC}} = 5$\,GeV together with the fit 
result. The slight deviations visible between fit and data are mostly due to the fact that the 
cylindrical geometry of LENA implies a direction-dependent angular resolution not accounted for by 
eq.~(\ref{eq:AngResFitFcn}). Therefore, our angular resolution $\sigma_{\alpha}$ essentially 
describes an average resolution.
The right side of figure \ref{fig:ResultInterAngle} presents the angular resolution we 
obtained from the fits for 1\,GeV bins of $E_{\mathrm{kin,MC}}$ between 1 and 10\,GeV. It decreases 
non-linearly from $(1.39 \pm 0.03)^{\circ}$ to $(0.30 \pm 0.01)^{\circ}$.

\begin{figure}[b!t]
  \centering
    \includegraphics[trim=0.3cm 0.0cm 0.6cm 0.7cm, clip=true, width=0.65\textwidth]
    {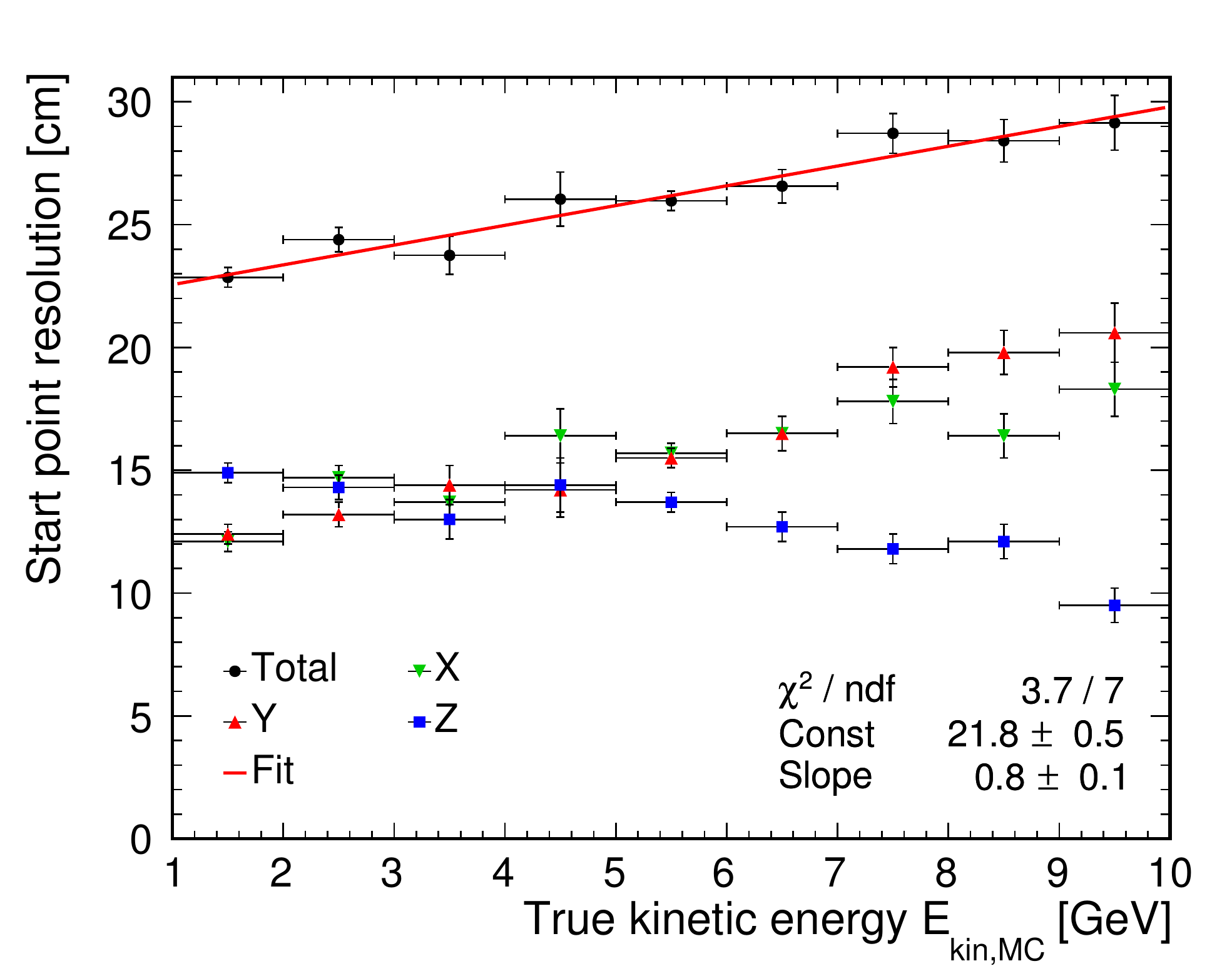}
  \caption{
Start point resolutions $\sigma_{s,x}$ (green tip down triangle), $\sigma_{s,y}$ (red tip up 
triangle) and $\sigma_{s,z}$ (blue square) along the detector coordinates $x$, $y$ and $z$ as a 
function of the true kinetic energy $E_{\mathrm{kin,MC}}$ in the range from 1 to 10\,GeV. Vertical 
error bars indicate the errors returned from the fits of eq.~(\ref{eq:StartPntCompFit}) to the 
distributions $u_{s,x}$, $u_{s,y}$ and $u_{s,z}$. The total resolution (black circle) was 
calculated according to eq.~(\ref{eq:TotalStartPntRes}) and fitted with a linear function.}
  \label{fig:ResultStartPointResoXYZ}
\end{figure}

\paragraph{Start point resolution} In a first step, we determined the resolution of $\mathbf{x}_s$ 
(for its reconstruction see \ref{sec:Analysis.AnaProcedure}) as a function of the $x$-, $y$- and 
$z$-direction in the LENA coordinate system, regarding the connecting vector $\mathbf{u}_s = \mathbf{x}_s - \hat{\mathbf{x}}_s$ 
between true and reconstructed start point. The resulting distributions of the components 
$u_{s,x}$, $u_{s,y}$, and $u_{s,z}$ exhibited a Gaussian distribution around zero on top of an 
essentially flat plateau explained further below. Consequently, we performed binned likelihood fits with 
the function
\begin{equation}
 F(u_{s,c}) = A \, \exp\left(-\frac{u_{s,c}^2}{2\sigma^2_{s,c}} \right) + B \, , \quad c = x,\, y,\,
z \, ,
\label{eq:StartPntCompFit}
\end{equation}
to extract the component-wise resolution illustrated in figure \ref{fig:ResultStartPointResoXYZ}.
We also show the total start point resolution 
\begin{equation}
 \sigma_{s,\mathrm{tot}} = \sqrt{\sigma_{s,x}^2 + \sigma_{s,y}^2 + \sigma_{s,z}^2} \, .
\label{eq:TotalStartPntRes}
\end{equation}
that follows a linear trend increasing with $(0.8\pm 0.1)\,\mathrm{cm} / \mathrm{GeV}$ and 
reaches $\sim30$\,cm at 10\,GeV. 

While the resolution is equal in all coordinates at lower energies, it becomes obvious above 6\,GeV 
that the uncertainty in $z$-direction decreases while the uncertainties in $x$- and $y$-direction increase.
This counter-intuitive finding arises from the preferentially vertical orientation of
fully-contained muon tracks that become increasingly aligned with the $z$-axis (symmetry axis) 
of the detector at higher energies (see figure \ref{fig:TrueDists} right) and from a difference in the start
point resolution lateral and parallel to the reconstructed track. Therefore, we studied the $\mathbf{x}_s$
resolution also relative to the track orientation, identifying a resolution parallel ($\sigma_{s,\mathrm{para}}$) 
and lateral ($\sigma_{s,\mathrm{lat}}$) to the reconstructed muon track based on the respective components
$u_{s,\mathrm{para}} = \mathbf{u}_s \cdot \hat{\mathbf{d}} / |\hat{\mathbf{d}}|$ and 
$u_{s,\mathrm{lat}}  = |\mathbf{u}_s \times \hat{\mathbf{d}}| / |\hat{\mathbf{d}}|$ of the connecting vector 
$\mathbf{u}_s$. We found that the distribution of $u_{s,\mathrm{para}}$ peaks around zero but exhibits
a long tail to negative values, i.e., a shift of $\hat{\mathbf{x}}_s$ along the track direction.
In the fits to determine $\sigma_{s,\mathrm{para}}$ this tail was approximated (within the fit range) 
by a constant term at the left side of the Gaussian describing the peak and yielding $\sigma_{s,\mathrm{para}}$.
This artifact is introduced by the current procedure to determine $\hat{\mathbf{x}}_s$ 
for cases where the reference point $\mathbf{x}_{\mathrm{ref}}$, which is projected onto the fitted line to find 
$\hat{\mathbf{x}}_s$, is outside of the selected event region. The systematic effect is also the cause for 
the two-sided tails in the distributions of $u_{s,x}$, $u_{s,y}$, and $u_{s,z}$ reported above. 
Since this issue is well understood and can be resolved in the future, we refrain from further investigation. 

\begin{figure}[b!t]
  \centering
  \begin{minipage}{0.48\textwidth}     
    \includegraphics[trim=0.1cm 0.1cm 0.0cm 0.1cm,clip=true,width=\textwidth]
      {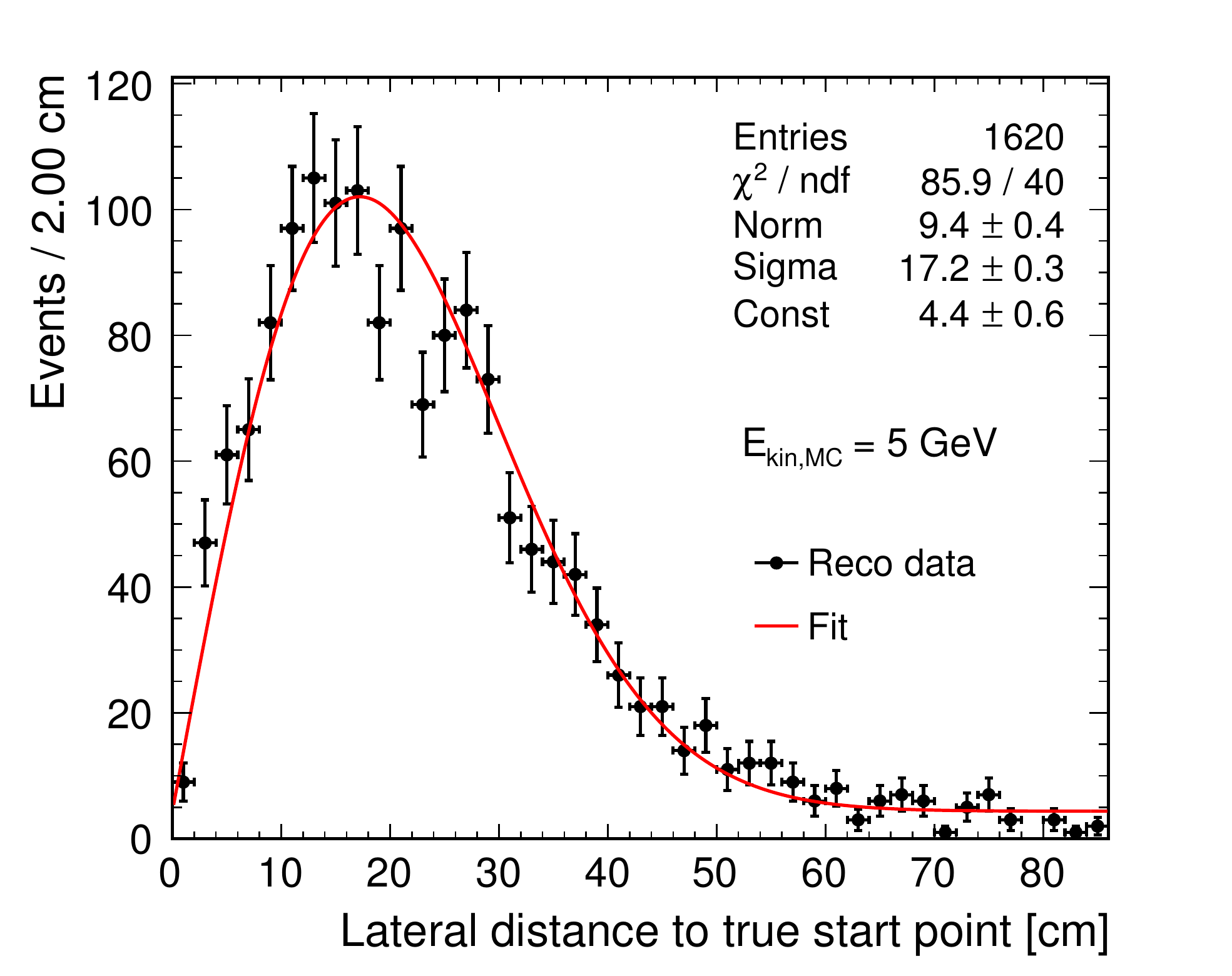}
  \end{minipage}
  ~
  \begin{minipage}{0.48\textwidth}
    \includegraphics[trim=0.1cm 0.1cm 0.0cm 0.1cm,clip=true,width=\textwidth]
      {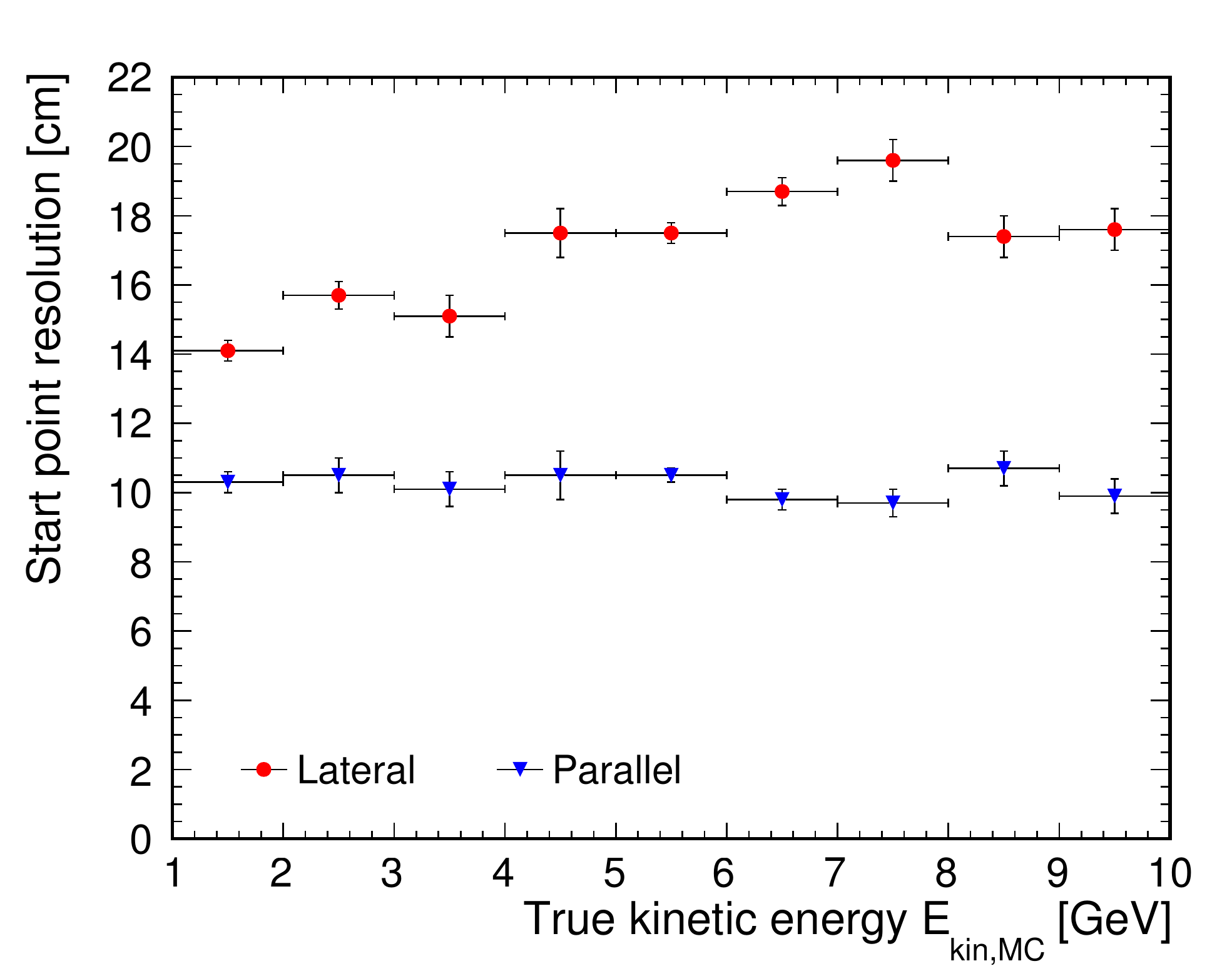}
  \end{minipage}
  \caption{
Left: Distribution of the lateral component $u_{s,\mathrm{lat}}$ of the connecting vector between 
true and reconstructed track start point for the high statistics sample at $E_{\mathrm{kin,MC}} = 
5\,\mathrm{GeV}$. It was fitted with eq.~(\ref{eq:StartPntLatFit}).
Right: Lateral (red circle) and parallel (blue triangles) start point resolutions 
$\sigma_{s,\mathrm{lat}}$ and  $\sigma_{s,\mathrm{para}}$ per 1\,GeV bin of the true muon kinetic 
energy $E_{\mathrm{kin,MC}}$ from 1 to 10\,GeV.}
  \label{fig:StartPntResLatPara}
\end{figure}

An example distribution for $u_{s,\mathrm{lat}}$ is shown on the left of figure 
\ref{fig:StartPntResLatPara}. To determine the resolution $\sigma_{s,\mathrm{lat}}$, we fitted it 
with the function
\begin{equation}
F(u_{s,\mathrm{lat}}) = u_{s,\mathrm{lat}} \, A \, \exp \left(-
\frac{u_{s,\mathrm{lat}}^2}{2\sigma_{s,\mathrm{lat}}^2} \right) + B \, ,
\label{eq:StartPntLatFit}
\end{equation}
where we included a constant term $B$ to approximate a tail to higher lateral distances. The right 
of figure \ref{fig:StartPntResLatPara} displays $\sigma_{s,\mathrm{para}}$ and 
$\sigma_{s,\mathrm{lat}}$ as a function of true kinetic energy.

Since we reconstructed start point $\hat{\mathbf{x}}_s$ by projecting the reference point 
$\mathbf{x}_{\mathrm{ref}}$ onto the three-dimensional track fit and $\mathbf{x}_{\mathrm{ref}}$ was
obtained by smearing the true start point $\mathbf{x}_s$ with random offsets of $\sigma = 10\,\mathrm{cm}$ 
per direction, the parallel resolution of the start point essentially inherits the uncertainty of 
$\mathbf{x}_{\mathrm{ref}}$. 

\paragraph{End point resolution} 
The analysis of the end point resolution $\sigma_e$ replicates that of $\sigma_s$. It is based on the 
connecting vector $\mathbf{u}_e = \mathbf{x}_e - \hat{\mathbf{x}}_e$ between the true end point 
$\mathbf{x}_e$ and the reconstructed point $\hat{\mathbf{x}}_e$. 

As mentioned in section \ref{sec:Analysis.AnaProcedure}, our analysis procedure took the exit point 
of the fitted line through the event region as reconstructed end point of the track. This definition 
crucially depends on the threshold condition for $\Gamma_{\mathrm{em}}(\mathbf{x})$ applied in the 
selection of the event region. In our case, this led to a systematic overestimation of the track
length beyond the true end point $\mathbf{x}_e$. The distribution of the parallel component 
$u_{e,\mathrm{para}}$ therefore showed a broad peak at negative values. A tail towards positive 
values of $u_{e,\mathrm{para}}$ was populated by events where the event region ``decayed'' 
into disconnected fragments, resulting in a diminished event region and too short tracks.

\begin{figure}[!tb]
  \centering
    \includegraphics[width=0.65\textwidth]{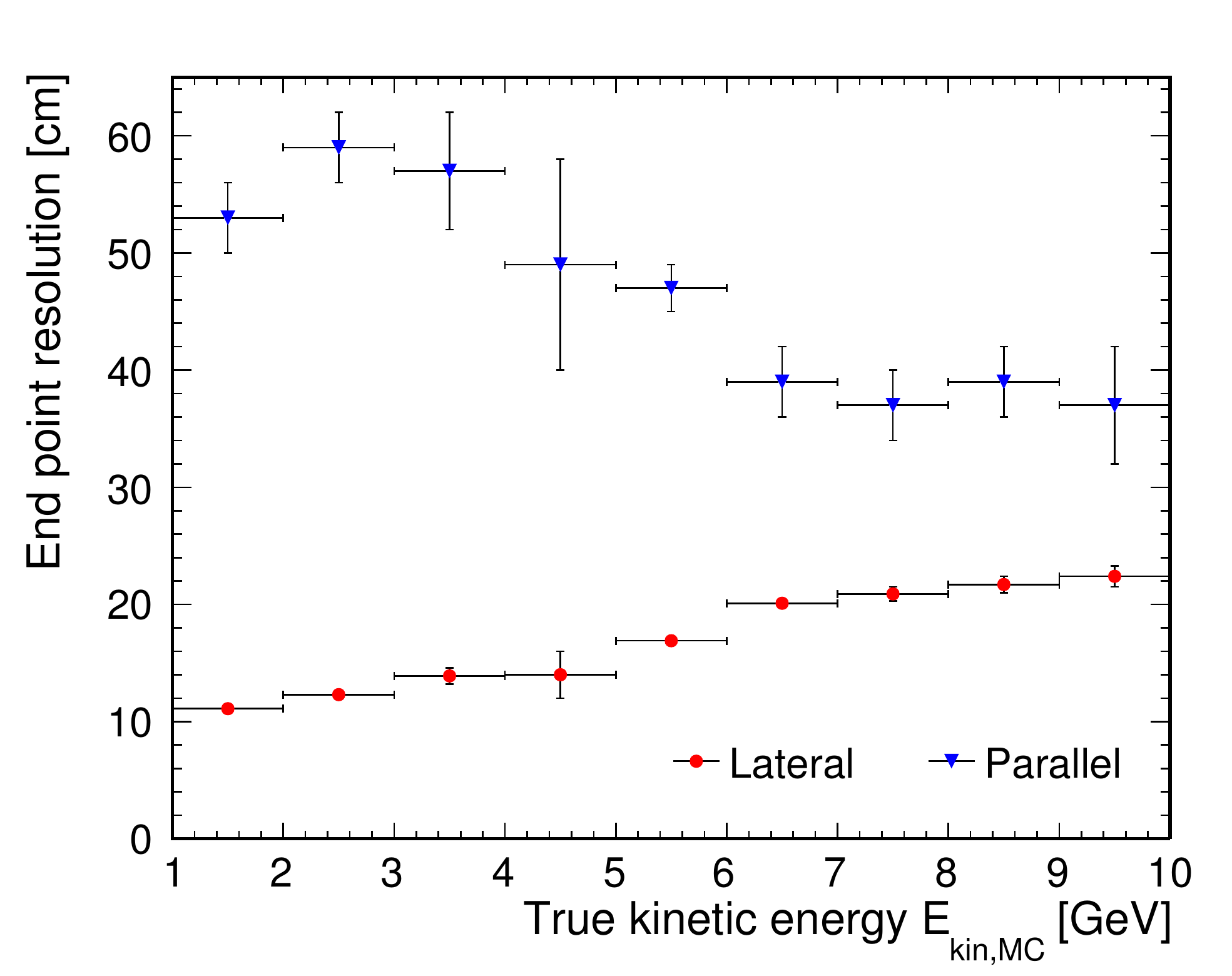}
    \caption{
Lateral (red circle) and parallel (blue triangles) end point resolutions $\sigma_{e,\mathrm{lat}}$ 
and  $\sigma_{e,\mathrm{para}}$ per 1\,GeV bin of the true muon kinetic energy 
$E_{\mathrm{kin,MC}}$ from 1 to 10\,GeV.}
  \label{fig:EndPntResLatPara}
\end{figure}

We fitted the distribution of $u_{e,\mathrm{para}}$ for a given energy range with the sum of two 
Gaussians, one describing the shifted peak and one describing the tail. The offset decreases almost 
linearly with increasing muon energy from 200\,cm at around 1\,GeV to about 120\,cm around 10\,GeV.
This systematic offset could either be corrected for by calibration or might be reduced by a more
sophisticated event region selection. Therefore, we regard only the standard deviation of the 
Gaussian describing the shifted peak as parallel resolution $\sigma_{e,\mathrm{para}}$. 
The energy-dependent result is shown in figure \ref{fig:EndPntResLatPara}, decreasing from 60\,cm to 
35\,cm with increasing energy. The figure also includes the lateral end point resolution 
$\sigma_{e,\mathrm{lat}}$ (determined equivalently to $\sigma_{s,\mathrm{lat}}$) that is not 
affected by the systematic parallel shift and ranges from 10\,cm to 22\,cm with increasing energy. 

\begin{figure}[!tb]
  \centering
    \includegraphics[width=0.80\textwidth]{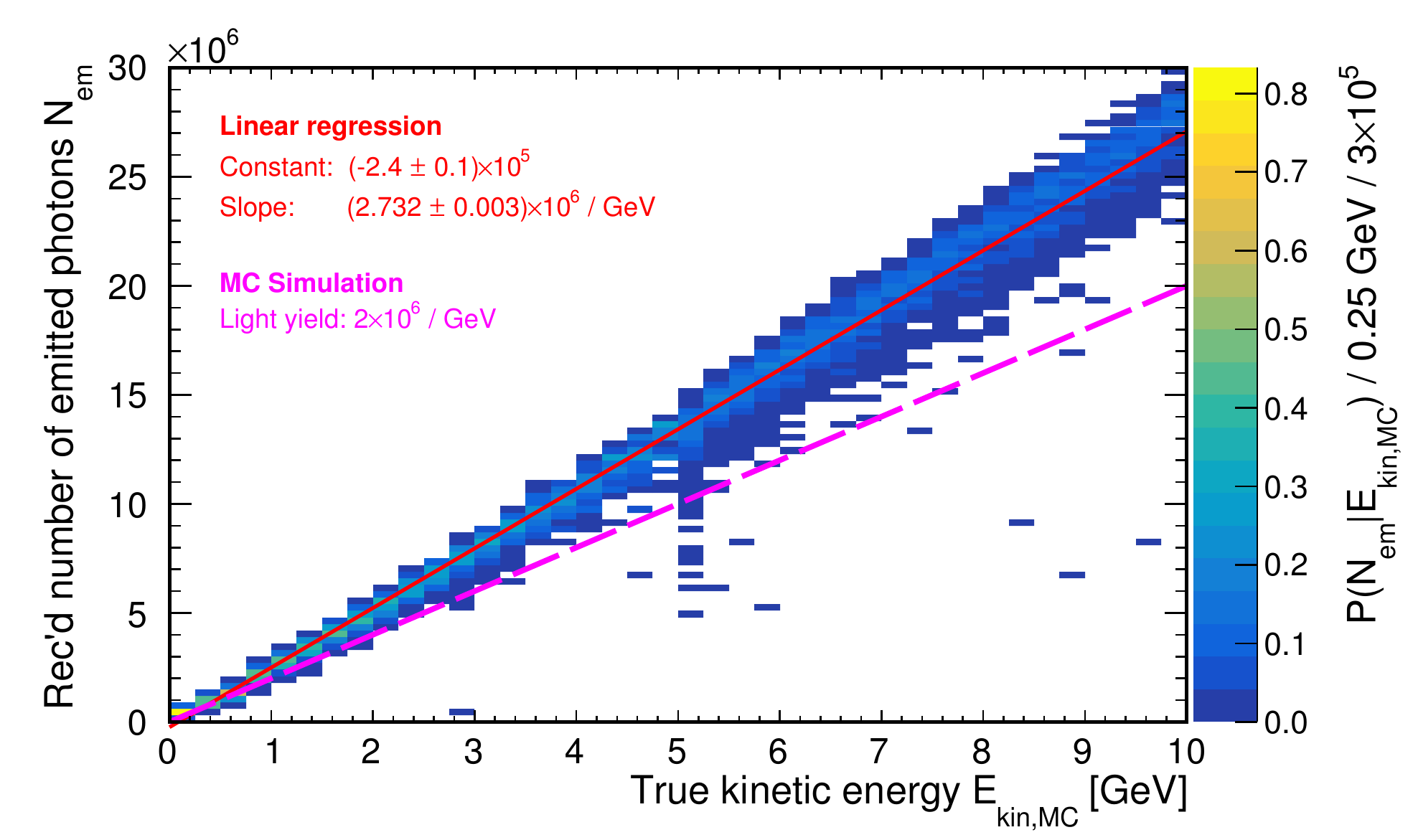}
    \caption{
Binned relative frequency distributions $P(\hat{N}_{\mathrm{em}}|E_{\mathrm{kin,MC}})$ as a 
function of the true kinetic energy $E_{\mathrm{kin,MC}}$. The distributions were created from the
sample of simulated muons shown in figure \ref{fig:TrueDists}. Note that the details at 5\,GeV
appear as a consequence of the high-statistics sample at this energy. The red (dashed magenta) line 
shows the result from a linear regression performed with the underlying data (the expectation from 
the MC settings).}
  \label{fig:ResultNumEmitPhotons}
\end{figure}

\paragraph{Total number of photon emissions / energy resolution}
The total number $\hat{N}_{\mathrm{em}}$ of photons emitted per event was obtained by the volume 
integral of $\Gamma_{\mathrm{em}}$ over the event region. Figure \ref{fig:ResultNumEmitPhotons} 
shows distributions of $\hat{N}_{\mathrm{em}}$ normalized per true energy bin and a line fit to the 
most frequent $\hat{N}_{\mathrm{em}}$ value for each true kinetic energy bin.
We find a light yield of $(2.732 \pm 
0.003)\times 10^6$\,GeV${}^{-1}$ that is about 37\% higher than the MC setting of 
$2\times10^6$\,GeV${}^{-1}$ and a negative value of $(-2.4 \pm 0.1) \times 10^5$ for the 
constant. The latter point suggests a loss of photons, e.g., due to the limitation of the volume 
integral to the event region or the actual escape of secondary particles from the LS volume.
The systematic increase of the reconstructed light yield is a direct consequence of neglecting 
scattered photons in the determination of the global detection efficiency according to 
eq.~(\ref{eq:LocalDefEff}). The resulting lower values of $\varepsilon(\mathbf{x})$ enter into the 
computation of $\Gamma_{\mathrm{em}}$ according to eq.~(\ref{eq:NumDensDistEmitted}) and yield
higher values compared to the expectation. However, this could be calibrated out based on MC.

\begin{figure}[!tb]
  \centering
    \includegraphics[width=0.80\textwidth]{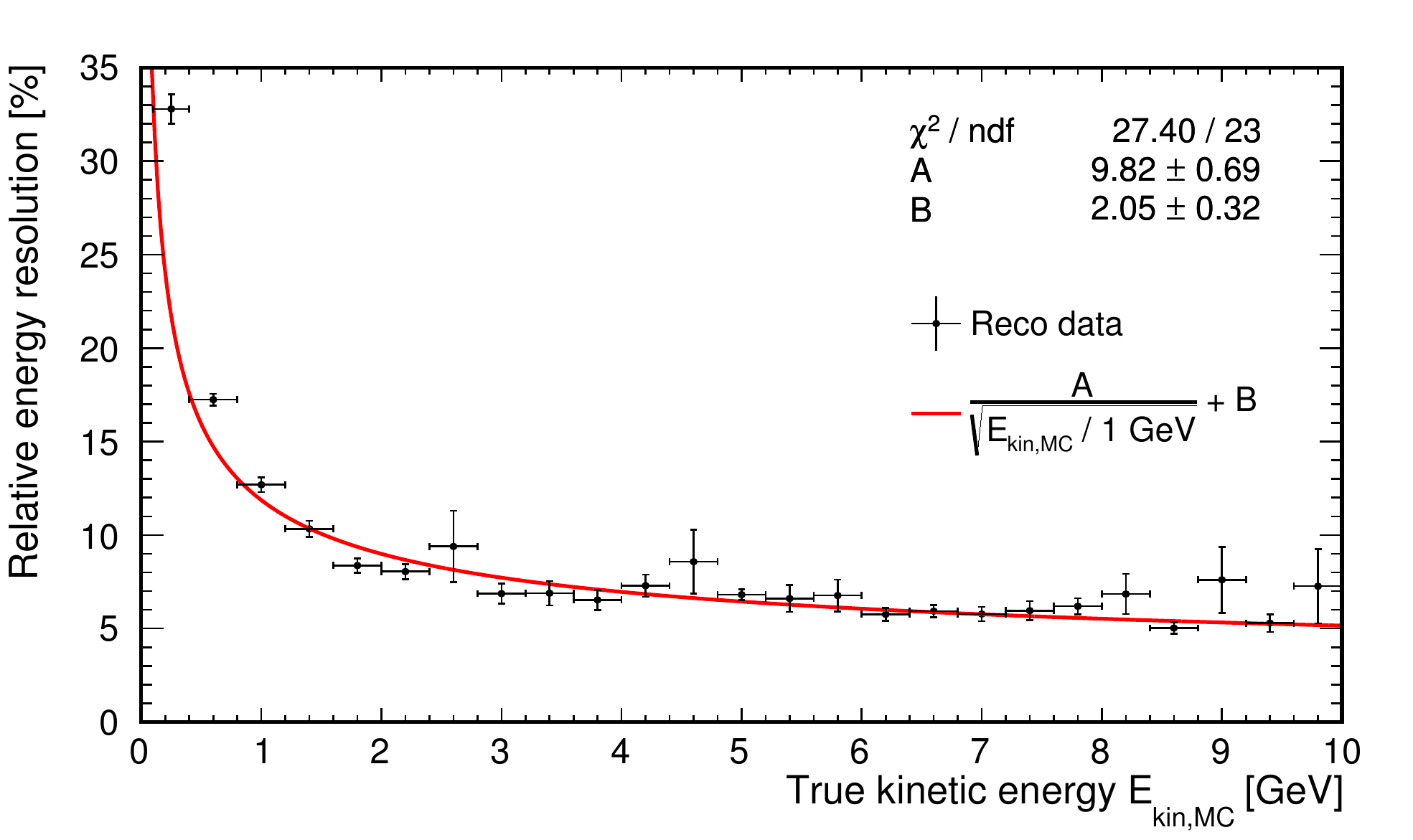}
    \caption{
Relative energy resolution as a function of the true kinetic energy of the stopping muons. The 
resolution was calculated as sample standard deviation over sample mean from the distribution of 
$\hat{N}_{\mathrm{em}}$ in each energy bin.}
  \label{fig:ResultEnergyResolution}
\end{figure}

Taking the non-Gaussian tails to lower values of $\hat{N}_{\mathrm{em}}$ into account, we estimated
the relative energy resolution $\sigma_{E} / E$ from the distribution of $\hat{N}_{\mathrm{em}}$
per energy bin as sample standard deviation over sample mean. The results for different energy 
ranges are shown in figure \ref{fig:ResultEnergyResolution}. They can be roughly described by the 
fitted function
\begin{equation}
 \frac{\sigma_{E}} {E} 
 \approx
  \frac{A}{\sqrt{E_{\mathrm{kin,MC}} / 1\,\mathrm{GeV}}} + B  = 
  \frac{(9.8 \pm 0.7)\%}{\sqrt{E_{\mathrm{kin,MC}} / 1\,\mathrm{GeV}}} + (2.0 \pm 0.3)\% \, .
\end{equation}

\paragraph{Remarks on reconstruction results below 1\,GeV} In the results presented above, the energy 
region below 1\,GeV is omitted for the angular, start and end point resolution. This is because the 
respective distributions exhibit large spreads as a consequence of failed track fits. For the latter we
identified two reasons: Firstly, the lower contrast in the reconstruction result $\Gamma_{\mathrm{em}}
(\mathbf{x})$ at lower energies combined with the finitely small extent of the reconstruction mesh cells 
lead to the selection of a quite compact event region without a distinct elongation in track direction. 
As a consequence, the line fit is dominated by random fluctuations of the region's shape. 
Secondly, the smearing of the reference parameter set with respect to the true values becomes significant 
relative to the overall track dimensions. This reduces the quality of the reconstruction and can 
occasionally lead to the ``decay'' of the true event region into several disconnected regions during the selection.
Since the algorithm always selected the (compact) sub-region closest to the reference point for further 
processing, the line fit is again dominated by random shape fluctuations as in the first case.
Therefore, requirements on the goodness of the reference parameters to achieve acceptable reconstruction
quality for $\Gamma_{\mathrm{em}}(\mathbf{x})$ have to be defined in future studies.

\paragraph{Comparison with other reconstruction methods} A direct comparison with 
``more conventional'' tracking algorithms is difficult, due to different detector 
geometries, different levels of detail in MC simulations, different resolution measures or 
the comparison between data and MC. Nevertheless, we summarize here a few results of other muon 
reconstruction methods:

A sophisticated method based on a likelihood fit was applied to the same 
LENA MC events of the sub-samples with [0.1,\,1]\,GeV and 0.5\,GeV~\cite{Hellgartner:2015_PhDThesis}, 
which we excluded from most of our studies. Compatible with our results at 1\,GeV, the angular
resolution reaches about $1.6^{\circ}$ at this energy and a fit to the obtained energy-dependent 
resolutions asymptotically approaches about $1.2^{\circ}$ for higher energies. The lateral resolution for
the start point of the muons and the relative energy resolution were found to be 2--5\,cm and $\sim1$\%,
respectively. A muon reconstruction for the spherical 20\,kt LS detector JUNO tested with events from a
a detailed MC simulation achieves a mean deviation between true and reconstructed angle of about 
$1.6^{\circ}$ over the entire detector~\cite{Genster:2018_ConeMethod}. For the resolution of the minimum 
track distance to the detector center, essentially describing the lateral resolution, a value of 20\,cm is reported.
A similar method has been used to track through-going muons in the much smaller, cylindrical Double 
Chooz detectors~\cite{Abe:2014_DCmuReco}, achieving a spatial resolution of 4\,cm per transverse direction
for tracks close to the center. Borexino found an angular resolution of $(2.44 \pm 0.19)^{\circ}$ and a
lateral track resolution of 30--40\,cm~\cite{Bellini:2011_BX_MuonNeutDet} for a sample of CNGS muons that 
crossed both OPERA~\cite{Agafonova:2015OPERA5thEvt} and the LS-filled inner vessel of the detector.

Although our method does not yet achieve the precision of muon reconstruction demonstrated
by Double Chooz, its first performance evaluation presented here nevertheless indicates the method's
competitiveness. This is even more so due to the method's advanced potentialities in terms of detailed 
event topology analyses.

\section{Outlook}
\label{sec:Outlook}

While we demonstrated that the topological reconstruction technique yields competitive results,
there is still considerable potential for improvements and extensions: First studies indicate that 
scattered photons can be largely removed as a contamination with the help of statistical methods. 
This not only saves computation time otherwise wasted on the useless processing of ``noise'', but 
also significantly increases the method's robustness and the final contrast of the result. 
Further necessary improvements include the proper handling of Cherenkov light as well as the 
integration and study of a more sophisticated optical model with a full set of wavelength-dependent 
optical properties and realistic detector signals (waveform-analysis).

Another aspect that needs to be addressed is computing time. Due to the reconstruction's 
computational complexity, the processing of a single muon event from our sample in section 
\ref{sec:Analysis.EventSample} took several hours on a single CPU, depending on the number of 
detected photon hits. This limited us to a final voxel size of 12.5\,cm although the amount of 
un-scattered light in a detector like LENA would in ideal cases lead to a resolution of a few 
centimeters. There is no reason to assume that the spatial resolution of the new method will fall 
short of the point-like vertex resolution of a few centimeters demonstrated by Borexino and other 
detectors for MeV event energies~\cite{Back:2012BxCalib}. Rigorous hit time data selection and parallelized 
computing, possibly with GPUs, are promising measures to considerably reduce the computing time. 
In first studies we already managed to reconstruct single muon events sufficiently well in a couple 
of minutes.

Preliminary results indicate that the resolution achievable with the topological reconstruction suffices 
to provide valuable information about the topology of MeV events traditionally considered as point-like. 
This could help to distinguish between single-site (electrons) and multi-site events (positrons, gammas) 
for background reduction purposes. However, a challenge here are the low photon hit counts because the 
method's iteration process is susceptible to statistical fluctuations. This makes a proper handling of 
scattered light crucial in this context.

Since the key ingredients of the method, the time matching of signals based on a reference point in 
space and time, the superposition of PDFs and the iterative procedure with self-generated prior 
information, are in no way tied to the technology of unsegmented, large-volume LS neutrino detectors, 
we see a number of other applications of the novel technique. The most obvious one is to use it for 
large water / ice Cherenkov detectors like Super-Kamiokande \cite{Fukuda:2002_SKDet} 
(Hyper-Kamiokande \cite{DiLodovico:2017HyperK}), KM3Net \cite{Adrian-Martinez:2016KM3Net} or 
IceCube \cite{Achterberg:2006IceCube}).\footnote{Note that the quality of the topological reconstruction 
for these detector types may be limited due to lower photon statistics compared to LS experiments.} 
Here, the Cherenkov angle can be utilized  to get additional information about the particles involved, 
but in turn makes the precomputed LUTs for detection efficiencies also direction-dependent. 
Water-based liquid scintillator (THEIA \cite{Alonso:2014_ASDC}) would give even more information because 
both light types could be used to provide complimentary information about each track. This would profit 
significantly from emerging technologies like large-area picosecond photodetectors (LAPPDs) \cite{Adams:2015LAPPDs}. 
Besides the timing, their main advantage is the small pixel size, which guarantees clearly defined 
single photon hits in contrast to a continuous waveform that is susceptible to complex pile-up effects 
from fast consecutive photon hits at large PMTs. 

To some extent, even ground based arrays looking at atmospheric showers could profit from the
reconstruction technique. Although they have instrumentation only on one side of the detection volume, 
our method should still be helpful to access the lateral air-shower profile, which would provide 
information on the primary particle inducing the shower.

Lastly, an application beyond particle physics is connected to particle beams that are employed in
medical physics or material science: Highly collimated beams (``pencil beams'') are used in proton 
or heavy-ion tumor therapy as well as in the analysis of materials based on X-ray fluorescence to 
resolve a sample's three-dimensional elemental distribution. Moreover, X-ray fluorescence allows to 
trace an incorporated, non-toxic pharmaceutical marker through a patient or subject \cite{BazalovaCarter:2015_XRayFluorescence}. 
The precise localization of the origins of X-rays with the corresponding wavelength then helps in 
diagnostics or understanding of the metabolism. If the patient / subject / material sample is surrounded 
by enough sensors to collect sufficient gamma- or X-rays from interactions of the beam within the medium 
or with the marker, the presented reconstruction method can be used to reconstruct the interaction density 
inside the medium (along the axis of the collimated beam). This could help to enhance the dose monitoring 
in particle therapies (prompt Gamma-Ray timing \cite{Golnik:2014_PromptGammaParticleTherapy, 
Hueso:2015_PromptGammaTimingTest}) or the precision of elemental distribution 
studies and marker tracing.

\section*{Summary}
\addcontentsline{toc}{section}{Summary}
\label{sec:Summary}

We have developed a novel reconstruction method for LS detectors that provides information on the 
topology of an extended event. The technique relies on the computation of the spatial probability density distribution 
regarding the origin of each detected scintillation photon. The computations take relevant effects 
of the photon production, propagation and detection inside an unsegmented LS detector into 
account. By combining the contributions of all detected photons in an iterative procedure, the method  
provides a three-dimensional map that represents the reconstructed distribution of 
scintillation photon emissions. This map not only reflects an event's topology but also provides 
access to the differential energy loss $\mathrm{d} E / \mathrm{d} x$. Both aspects are important 
fundamentals to establish powerful particle identification and background rejection techniques for 
the LS technology.

Using a state-of-the-art Geant4 simulation of the multi-kiloton LS detector LENA, we tested
the reconstruction performance with track-like topologies from fully-contained, single 
muons that had kinetic energies up to 10\,GeV. Since the topological reconstruction method requires 
a second stage to extract descriptive event parameters from the output distribution, we have 
also started to develop a set of analyses for this purpose. 

Our first analysis outcomes brought up a few critical items. In particular, the accuracy and 
robustness of the start / end point determination for our fully-contained tracks are 
currently diminished by systematic effects related to the current analysis procedure. This mostly 
lead to energy-dependent resolutions, especially parallel to the reconstructed track direction, that 
are momentarily inferior to reasonable expectations. However, the outstanding issues in the 
analysis stage can be resolved in the future. The start (end) point resolution lateral to 
the reconstructed track, which is less affected by the systematic effects, is below 22\,cm (24\,cm) 
in the kinetic energy region between 1 and 10\,GeV. 

For the angular resolution of the simulated muon tracks we found a decrease from 
$(1.39\pm0.03)^{\circ}$ to $(0.30\pm0.01)^{\circ}$ between 1 and 10\,GeV. Moreover, our results 
concerning the reconstructed total number of scintillation photon emissions exhibit the almost 
linear dependence on the true kinetic energy expected for muons. Therefore, leaving the resolvable 
analysis issues aside, this first performance evaluation shows that the novel method is indeed fully 
competitive to track reconstruction codes employed in earlier LS neutrino experiments~\cite{
Bellini:2011_BX_MuonNeutDet, Abe:2014_DCmuReco}. Moreover, the demonstrated power to 
resolve topological features of an event exhibits the potential to complement or surpass existing 
reconstruction methods for LS detectors. Since the underlying key concepts of the topological 
reconstruction method are not tied to the LS technology, the method can in principle be applied for 
other detector types or even in other fields of physics, like medical physics or material science.

\section*{Acknowledgements}
\addcontentsline{toc}{section}{Acknowledgements}
\label{sec:Acknowledgements}

We thank Randolph M\"ollenberg for the strong support of our work regarding the LENA Geant4 detector simulation.
For the production of our event sample we benefited from access to the Batch Infrastructure Resource 
at DESY (BIRD) in Hamburg and to the Finnish Grid Infrastructure (FGI). This work was funded by the 
LAGUNA-LBNO design study (Grant Agreement No. 284518) in the context of the 7th 
Framework Programme for Research and Technological Development of the European Union and by the Deutsche 
Forschungsgemeinschaft with the PRISMA cluster of excellence at the Johannes Gutenberg-University Mainz.

\addcontentsline{toc}{section}{Bibliography}
\bibliographystyle{JHEP}
\bibliography{bibliography}

\end{document}